\documentclass[11pt,amsfonts,floatfix,superscriptaddress,longbibliography,footinbib]{revtex4-2}
\usepackage[utf8]{inputenc}
\usepackage{amsmath}    % need for subequations
\usepackage{amssymb}
\usepackage{mathbbol}
\usepackage{graphicx}   % need for figures
\usepackage{verbatim}   % useful for program listings
\usepackage{subfigure}  % use for side-by-side figures
\usepackage{hyperref} 
\usepackage{scalerel}
\usepackage{cancel}
\usepackage{soul}
\usepackage[normalem]{ulem}
\usepackage[dvipsnames]{xcolor}
\usepackage{siunitx}
\usepackage{physics}
\usepackage{enumerate} % http://ctan.org/pkg/enumerate
\usepackage{empheq}

\begin{document}

\title{Semiclassical entanglement entropy for spin-field interaction}

\author{Matheus V. Scherer}
\affiliation{Departamento de F\'{\i}sica, Universidade Federal do Paran\'a, 81531-990, Curitiba, PR, Brazil}

\author{Lea F. Santos}
\affiliation{Department of Physics, University of Connecticut, Storrs, Connecticut 06269, USA}

\author{Alexandre D. Ribeiro}
\affiliation{Departamento de F\'{\i}sica, Universidade Federal do Paran\'a, 81531-990, Curitiba, PR, Brazil}
\affiliation{Department of Physics, University of Connecticut, Storrs, Connecticut 06269, USA}

%%%%%%%%%%%%%%%%%%%%
%%%%%%%%%%%%%%%%%%%%
\begin{abstract}
   We study a general bipartite quantum system consisting of a spin interacting with a bosonic field, with the initial state prepared as the product of a spin coherent state and a canonical coherent state. Our goal is to develop a semiclassical framework to describe the entanglement dynamics between these two subsystems. Using appropriate approximations, we derive a semiclassical expression for the entanglement entropy that depends exclusively on the trajectories of the underlying classical description. By analytically extending the classical phase space into the complex domain, we identify additional complex trajectories that significantly improve the accuracy of the semiclassical description. The inclusion of these complex trajectories allows us to capture the entanglement dynamics with remarkable precision, even well beyond the Ehrenfest time. The approach is illustrated with a representative example, where the role of real and complex trajectories in reproducing the quantum entanglement entropy is explicitly demonstrated.
\end{abstract}

\maketitle

%%%%%%%%%%%%%%%%%%%%
%%%%%%%%%%%%%%%%%%%%
\section{Introduction}
\label{Intro}

Consider a quantum bipartite system composed of two parts, $A$ and $B$. A pure state $|\psi\rangle$ is said to be {\em entangled} if it cannot be factorized as $|\psi\rangle = |\psi_A\rangle \otimes |\psi_B\rangle$, with $|\psi_A\rangle$ and $|\psi_B\rangle$ referring to subsystems $A$ and $B$, respectively. Otherwise, $|\psi\rangle$ is termed {\em separable}. This fundamental notion of entanglement, introduced by Schr\"odinger~\cite{sch1935} nearly a century ago, became central to understanding the unusual correlations discussed in the celebrated Einstein–Podolsky–Rosen thought experiment~\cite{EPR}. Since those early debates, entanglement has often been regarded as a distinctly quantum feature, with no classical counterpart. Over the years, the concept has evolved to contemplate mixed and multipartite states, having reached a prominent role in problems involving quantum information processing~\cite{chuang,amicoRMP,horodeckiRMP}. More recently, its scope has expanded far beyond foundational discussions. Today, entanglement plays a key role in high-energy and gravitational physics~\cite{wittenRMP,nishiokaRMP,maldacena2003,qi,bianchi2014}, the study of thermalization in isolated quantum systems~\cite{calabrese2005,lea2012,rigolAP} and many approaches related to the field of quantum chaos and many-body dynamics~\cite{kyoko,ghose2004,Bandyopadhyay2002,Santos2002,lea2004,Santos2005,angelo2004,jaquod2004,novaes2005,prosen2005,Monasterio2005,matzkin2006,matzkin2006l,ribeiro2010,bonanca2011,ribeiro2012,casati2012,berenstein2016,neill2016,pattanayak2017,zheng2017,ribeiro2019,quach2019,ghose2019,pappalardi2020,ribeiro2021,pappalardi2024}.

In this work, we focus on entanglement in the context of the quantum–classical transition. One of the most emblematic results in this area is the link between the rate of entanglement growth and the Lyapunov exponents of the underlying classical dynamics, as discussed in the seminal work by Furuya, Nemes, and Pellegrino~\cite{kyoko}. This implies that when one initializes a localized separable quantum state in phase space, its entanglement increases in time at a rate determined by the degree of classical chaoticity of the region it occupies. This establishes a direct correspondence between chaos and entanglement generation. A promising route to understanding this relationship consists of introducing semiclassical entropy formulas that reveal which classical ingredients govern entanglement dynamics. This approach has been successfully implemented for systems of two interacting particles in the canonical (position and momentum) phase space~\cite{jaquod2004,ribeiro2010,ribeiro2019} and for coupled spins~\cite{ribeiro2012,ribeiro2021}. In both cases, the growth of entanglement is shown to be dictated by the stability of the classical trajectories, confirming the chaos-entanglement correspondence.

A key advance of these works was the use of coherent-state path-integral techniques, which accommodate an analytic continuation of the classical phase space into the complex domain. The inclusion of complex trajectories extends the validity of the semiclassical description well beyond the Ehrenfest time, allowing for accurate long-time predictions of entanglement entropy~\cite{ribeiro2019,ribeiro2021}. Building on this program, we now turn to a distinct system of broad physical relevance: a spin interacting with a bosonic field. This setup can be viewed as a minimal hybrid system combining a finite-dimensional spin degree of freedom with an infinite-dimensional canonical one. The derivations involve nontrivial modifications over previous studies and yield new insights into the role of complex trajectories in spin-field entanglement. Our analysis also provides a more detailed and systematic presentation of the semiclassical approach, making it broadly applicable to other hybrid quantum systems.

The paper is organized as follows. In Sec.~\ref{PCsec}, we review the essential background needed to
understand our calculation, including the properties of spin and canonical coherent states, and the derivation of semiclassical expressions for the forward and backward propagator. In Sec.~\ref{SEsec}, we derive the semiclassical expression for the entanglement entropy. This is illustrated in Sec.~\ref{Esec} with a simple but instructive example, where the role of real and complex trajectories is explicitly analyzed. We conclude in Sec.~\ref{FRsec} with a discussion of our main findings and possible directions for future work.

%%%%%%%%%%%%%%%%%%%%
%%%%%%%%%%%%%%%%%%%%
\section{Preliminary concepts}
\label{PCsec}

Forward $K_+^{\mathrm{q}}$ and backward $K_-^{\mathrm{q}}$ quantum propagators are the elementary quantities used in the present paper to study how the entanglement of pure bipartite states evolves from the initial time $t'=0$ to a final time $t''=T$ under a time-independent Hamiltonian $\hat{H}$. In the representation given by the product of canonical $|z\rangle$ and spin $|s\rangle$ coherent states, that is, $|z,s\rangle\equiv|z\rangle\otimes|s\rangle$, the propagators take the form
\begin{equation}
K_\pm^{\mathrm{q}} 
(z_{\mbox{\scriptsize b}}, s_{\mbox{\scriptsize b}}, z_{\mbox{\scriptsize k}}, s_{\mbox{\scriptsize k}}, T) \equiv
\langle  z_{\mbox{\scriptsize b}}, s_{\mbox{\scriptsize b}}|
\, \mbox{e}^{\mp i\hat{H}T/\hbar} \,|  z_{\mbox{\scriptsize k}} , s_{\mbox{\scriptsize k}}\rangle . 
\label{Kq}
\end{equation}

In what follows, we discuss some of the key ingredients underlying this expression. Section~\ref{CSsec} reviews the main properties of the coherent-state basis~$\{|z,s\rangle\}$. For a detailed treatment of this representation, we refer the reader to the classic references on the subject~\cite{klauderb, perelomov, gilmore,gazeau}. Since our focus is on the semiclassical regime of entanglement, Sec.~\ref{SPsec} introduces the semiclassical approximation to Eq.~\eqref{Kq}. While the {\em forward} propagator has been extensively studied in the context of coherent states over the past decades~\cite{scsp1, scsp2, scsp3, aguiar01, ribeiro2004, garg1, sscsp1, sscsp2, sscsp3, sscsp4, sscsp5, ribeiro2006, garg2, thiago}, the {\em backward} propagator has only more recently attracted attention~\cite{ribeiro2010, ribeiro2012, ribeiro2019, ribeiro2021, ribeiro2017}. To our knowledge, a semiclassical expression for $K_-^{\mathrm{q}}$ in the joint product basis $|z,s\rangle$ has not been considered before. One of our main contributions is to fill this gap by deriving such an expression, combining insights from Ref.~\cite{ribeiro2006}, where the approximation of $K_+^{\mathrm{q}}$ was found, as well as Refs.~\cite{ribeiro2010, ribeiro2012}, where the backward semiclassical propagator was achieved for {\em exclusively} canonical~\cite{ribeiro2010} and spin~\cite{ribeiro2012} bases.

%%%%%%%%%%%%%%%%%%%%
%%%%%%%%%%%%%%%%%%%%
\subsection{Coherent states}
\label{CSsec}

We consider a bipartite system with Hilbert space $\mathcal{H} = \mathcal{H}_z \otimes \mathcal{H}_s$, where $\mathcal{H}_z$ describes a canonical infinite-dimensional degree of freedom and $\mathcal{H}_s$ corresponds to a spin-$j$ system of dimension $2j+1$. We define the canonical coherent-state basis $\{|z\rangle\}$ and the spin coherent-state basis $\{|s\rangle\}$. In terms of the Fock basis $\{|n_z\rangle\}$ and the basis $\{|m_s\rangle\}$, composed by the eigenstates of $\hat{J}_{{\mathrm{z}}}$, with the spin operator given by $\hat{\mathbf{J}}=(\hat{J}_{{\mathrm{x}}},\hat{J}_{{\mathrm{y}}},\hat{J}_{{\mathrm{z}}})$, we have
\begin{equation}
|z \rangle = \mathcal{N}_z \, \mbox{e}^{z \hat{a}^{\dagger}} |n_z=0\rangle 
\quad \mathrm{and} \quad 
|s \rangle = \mathcal{N}_s \, \mbox{e}^{s \hat{J}_+}  | m_s=-j \rangle ,
\label{CSdef}
\end{equation}
where $\mathcal{N}_z \equiv \mbox{e}^{-\frac{1}{2} |z|^2}$ and $\mathcal{N}_s \equiv (1+ |s|^2 )^{-j}$ ensure normalization, $\hat{a}^{\dagger}$ is the bosonic creation operator, $|n_z=0 \rangle$ is the Fock vacuum state, $\hat{J}_+\equiv\hat{J}_{{\mathrm{x}}}+i\hat{J}_{{\mathrm{y}}}$ is the raising spin operator, and $| m_s=-j \rangle$ is the extremal eigenstate of $\hat{J}_{{\mathrm{z}}}$ with the lowest eigenvalue. The complex numbers $z$ and $s$ are the labels of the coherent states, and their meaning will be clarified later.

The identity operators for the canonical and spin bases are given, respectively, by
\begin{equation}
\hat{\mathrm{I}}_z  \equiv
\int \frac{ |z \rangle \langle z| \,\mbox{d}z^{\text{\tiny R}} \, \mbox{d}z^{\text{\tiny I}} }{\pi}
\quad \mathrm{and} \quad 
\hat{\mathrm{I}}_s  \equiv
\frac{2j+1}{\pi} \int \frac{|s \rangle \langle s| \, \mbox{d}s^{\text{\tiny R}} \, \mbox{d}s^{\text{\tiny I}} }
{\left( 1 + |s |^2 \right)^{2}} , 
\label{IdOp} 
\end{equation}
where $z^{\text{\tiny R}}$ ($z^{\text{\tiny I}}$) indicates the real (imaginary) part of the complex label $z$, and equivalently for $s$. In the above equations, the first integration should run over the whole complex plan $z$, while the second one runs over the complex plan $s$. 

Both bases are overcomplete, as their elements are not mutually orthogonal. Indeed, the overlaps between any two basis states read  
\begin{equation}
\langle z_{\mbox{\scriptsize b}} | z_{\mbox{\scriptsize k}} \rangle = 
\mathcal{N}_{z_{\mbox{\tiny b}}}\mathcal{N}_{z_{\mbox{\tiny k}}} 
\mbox{e}^{z_{\mbox{\tiny b}}^* z_{\mbox{\tiny k}}}
\quad \mathrm{and} \quad 
\langle s_{\mbox{\scriptsize b}} | s_{\mbox{\scriptsize k}} \rangle = 
\mathcal{N}_{s_{\mbox{\tiny b}}}\mathcal{N}_{s_{\mbox{\tiny k}}} 
\left( 1 + s_{\mbox{\scriptsize b}}^* s_{\mbox{\scriptsize k}} \right)^{2j}.
\label{overlap}
\end{equation}

\subsubsection{Canonical coherent states}

The complex number $z$ that labels the canonical coherent state $|z\rangle$ can be parametrized as
\begin{equation}
z = \frac{1}{\sqrt{2}}\left(\frac{q}{b} + i\frac{p}{c}\right), 
\label{zs}
\end{equation}
where $bc = \hbar$ and $(q,p,b,c)$ are convenient new parameters. 

Using $\hat{a} |z\rangle = z |z\rangle$ and the relations for the position $\hat{q}$ and momentum $\hat{p}$ operators,
\begin{equation}
\hat{q} = \frac{b}{\sqrt2}\left( \hat{a}^\dagger + \hat{a}\right) 
\quad\mathrm{and}\quad
\hat{p} = \frac{ic}{\sqrt2}\left( \hat{a}^\dagger - \hat{a}\right), 
\end{equation}
we obtain the expectation values and the variances of position and momentum, 
\begin{equation}
\begin{aligned}
&\langle \hat{q} \rangle_{z} = q, \quad
\Delta q^2 \equiv \langle \hat{q}^2 \rangle_{z} - \langle \hat{q} \rangle_{z}^2 
= b^2/2, \\
&\langle \hat{p} \rangle_{z} = p, \quad
\Delta p^2 \equiv \langle \hat{p}^2 \rangle_{z} - \langle \hat{p} \rangle_{z}^2
= c^2/2. \\
\end{aligned}
\label{zcalc}
\end{equation}
Thus, $|z\rangle$ saturates the Heisenberg uncertainty principle and represents the most classical states of the canonical degree of freedom.

\subsubsection{Spin coherent states}

The spin label $s$ that labels the spin coherent state $|s\rangle$ can be expressed in terms of spherical coordinates $(\theta,\phi)$ via the projection
\begin{equation}
s = \cot(\theta/2)(\cos\phi - i\sin\phi),
\label{zsS}
\end{equation}
where $0\le\theta\le\pi$ and $0\le\phi<2\pi$.

To understand the parametrization of $|s\rangle$, 
we define $\mathbf{n} = (\sin\theta\cos\phi, \sin\theta\sin\phi, \cos\theta)$, so that $\mathbf{n}$ is a versor spanning a sphere of radius one, assuming that $\theta=0$ ($\theta=\pi$) corresponds to its north (south) pole.
 Then, let the $s$ complex plane divide the sphere into north and south hemispheres, with the real and the imaginary axes pointing to $\phi=0$ and $\phi=3\pi/2$, respectively. In this way, we can perform a stereographical projection from the north pole, linking the sphere point $(\theta,\phi)$ with the coordinates $(s^{\text{\tiny R}},s^{\text{\tiny I}})$ of the complex plane, as represented in Eq.~\eqref{zsS}. Now, we calculate
\begin{equation}
\Delta J_l^2 \equiv \langle \hat{J}_l^2 \rangle_{s} - 
\langle \hat{J}_l\rangle_{s}^2 
= \frac{j}{2}(1-n_l^2), \qquad
\langle \big\{ \hat{J}_r - \langle\hat{J}_r \rangle, \hat{J}_l - 
\langle\hat{J}_l\rangle \big\} \rangle_s 
= -j \, n_r \, n_l,
\end{equation}
and $\langle \hat{\mathbf{J}} \rangle_{s} = j\mathbf{n}$, where $r\neq l$ can assume $\mathrm{x}$, $\mathrm{y}$, or $\mathrm{z}$. The operation $\{\star,\ast \}$ stands for anticommutator. With these results, we can conclude that the spin coherent state has the most well-localized representation in the unit sphere, as its uncertainties saturate the Heisenberg relation $\langle \Delta \hat{A}^2\rangle  \langle \Delta \hat{B}^2\rangle \ge \frac{1}{4}|\langle [\hat A,\hat B]\rangle|^2+ \frac{1}{4}|\langle \{\Delta\hat A,\Delta\hat B\}\rangle|^2$, where the operators $\hat{A}$ and $\hat{B}$ can be $\hat{J}_{{\mathrm{x}}}$, $\hat{J}_{{\mathrm{y}}}$, and $\hat{J}_{{\mathrm{z}}}$. In this sense, they can also be seen as the quantum states whose quantum description better approaches a classical angular momentum.

The canonical and spin coherent states thus both represent the most classical-like states in their respective spaces. Their use provides a natural framework for semiclassical approximations, justifying their role in the construction of our propagators and, ultimately, in the semiclassical entanglement entropy.

%%%%%%%%%%%%%%%%%%%%
%%%%%%%%%%%%%%%%%%%%
\subsection{Semiclassical forward and backward propagators}
\label{SPsec}

In this section, we will briefly outline the derivation of the semiclassical expressions $K^{\mathrm{sc}}_\pm$ for the quantum propagators defined in Eq.~\eqref{Kq}. Our approach follows the methods developed in Refs.~\cite{aguiar01, ribeiro2006, ribeiro2010, ribeiro2012}, where related problems were investigated.

%%%%%%%%%%%%%%%%%%%%
%%%%%%%%%%%%%%%%%%%%
\subsubsection{Path integral formalism}

To apply a semiclassical approximation to Eq.~\eqref{Kq}, we begin with its path-integral representation. This can be obtained by dividing the total propagation time $T$ into $N$ infinitesimal steps of duration $\epsilon$ and inserting identities~\eqref{IdOp} between successive short-time evolution operators $\hat{U}^{\pm}_{\epsilon}\equiv\mbox{e}^{\mp i\hat{H} \epsilon/\hbar}$. The resulting expression is
\begin{equation}
\begin{aligned}
K_\pm^{\mathrm{q}} &= 
\lim_{N\to\infty}
\int \left[
\prod_{\tau=1}^{N-1}
\left( \frac{2j+1}{\pi^2}\right) 
\frac{\mbox{d}z_{\tau}^{\mbox{\tiny R}} \, \mbox{d}z_{\tau}^{\mbox{\tiny I}} \, 
\mbox{d}s_{\tau}^{\mbox{\tiny R}} \, \mbox{d}s_{\tau}^{\mbox{\tiny I}}}{(1+|s_\tau|^2)^2}
\right] 
\\&\quad\quad\times
\langle  z_{\mbox{\scriptsize b}}, s_{\mbox{\scriptsize b}} | \, \hat{U}^{\pm}_{\epsilon} \, |  z_{1} , s_{1}\rangle
\left[ \prod_{\tau=1}^{N-2} 
\langle  z_{\tau} , s_{\tau} | \, \hat{U}^{\pm}_{\epsilon} \, |  z_{\tau+1} , s_{\tau+1}\rangle
\right]
\langle  z_{N-1} , s_{N-1}| \, \hat{U}^{\pm}_{\epsilon} \,|  z_{\mbox{\scriptsize k}} , s_{\mbox{\scriptsize k}}\rangle .
\end{aligned}
\label{KqPI}
\end{equation}
The interpretation of this equation is similar to that of the most-known propagator in the position representation. Indeed, in the space consisting of the integration variables, all paths connecting the ket labels ($z_{\mbox{\scriptsize k}} , s_{\mbox{\scriptsize k}}$) to the bra labels ($z_{\mbox{\scriptsize b}} , s_{\mbox{\scriptsize b}}$), from initial to final times, produce a complex contribution to the integral, and the exact quantum propagators $K_\pm^{\mathrm{q}}$ are obtained by summing over all of them.

Carrying out this procedure exactly is intractable in general. Progress is possible, however, in the semiclassical limit, which is formally achieved by assuming the asymptotic limit $\hbar\to0$ and $j\to\infty$, with the product $j\,\hbar$ held fixed. To perform this approximation, we first reinterpret Eq.~\eqref{KqPI} as a line integral in a multidimensional space. To understand this point, consider, for simplicity, only the variables $z_1^{\mbox{\tiny R}} $ and $z_1^{\mbox{\tiny I}}$. Although the original prescription in Eq.~\eqref{IdOp} imposes an integration over the entire plane $z_1$, if we define the new variables $v_{1\mbox{\tiny{A}}} \equiv z_1^{\mbox{\tiny R}} -i z_1^{\mbox{\tiny I}}$ and $u_{1\mbox{\tiny{A}}} \equiv z_1^{\mbox{\tiny R}} +i z_1^{\mbox{\tiny I}}$, that surface integral becomes equivalent to line integrals in the complex planes $v_{1\mbox{\tiny{A}}}$ and $u_{1\mbox{\tiny{A}}}$, now considered as independent variables. This equivalence is achieved by imposing that the line integral on $v_{1\mbox{\tiny{A}}}$ ($u_{1\mbox{\tiny{A}}}$) be along the real (imaginary) axis. Next, extending this construction to the other variables of the integral in~\eqref{KqPI} and calculating the appropriate Jacobian, we get
\begin{equation}
K_\pm^{\mathrm{q}} 
(z_{\mbox{\scriptsize b}}, s_{\mbox{\scriptsize b}}, z_{\mbox{\scriptsize k}}, s_{\mbox{\scriptsize k}}, T)
=
\lim_{N\to\infty} \int \left[\prod_{\tau=1}^{N-1}
\left( \frac{2j+1}{-4\pi^2}\right) 
\frac{\mbox{d} u_{\tau\mbox{\tiny{A}}} \, \mbox{d} v_{\tau\mbox{\tiny{A}}} \, 
\mbox{d} u_{\tau\mbox{\tiny{B}}} \, \mbox{d} v_{\tau\mbox{\tiny{B}}} }
{(1+ u_{\tau\mbox{\tiny{B}}} v_{\tau\mbox{\tiny{B}}})^2}\right]~\mbox{e}^{\mathcal{S}},
\label{KqLI}
\end{equation}
where $\mbox{e}^\mathcal{S}$ represents the second line of Eq.~\eqref{KqPI} expressed in terms of the new variables
\begin{equation}
v_{\tau\mbox{\tiny{A}}} \equiv z_\tau^{\mbox{\tiny R}} -i z_\tau^{\mbox{\tiny I}}, \quad 
u_{\tau\mbox{\tiny{A}}} \equiv z_\tau^{\mbox{\tiny R}} +i z_\tau^{\mbox{\tiny I}}, \quad
v_{\tau\mbox{\tiny{B}}} \equiv s_\tau^{\mbox{\tiny R}} -i s_\tau^{\mbox{\tiny I}}, \quad\mathrm{and}\quad
u_{\tau\mbox{\tiny{B}}} \equiv s_\tau^{\mbox{\tiny R}} +i s_\tau^{\mbox{\tiny I}} ,
\label{nv}
\end{equation}
for $\tau=1,\ldots,N-1$. Notice that, at this point, we already refer to the canonical degree of freedom as part A of the system and the spin variable corresponds to part B. With this procedure, the path integral formulation for $K_\pm^{\mathrm{q}}$ can be understood as a $4N$-dimensional complex line integral~\eqref{KqLI}.

%%%%%%%%%%%%%%%%%%%%
%%%%%%%%%%%%%%%%%%%%
\subsubsection{Complex trajectories}

The line integral in Eq.~\eqref{KqLI} can be evaluated using the saddle-point method~\cite{bleistein}. Instead of considering the contribution of all points belonging to the paths appearing in Eq.~\eqref{KqLI}, the method focuses exclusively on the critical points, around which the integral acquires its dominant contributions. In the semiclassical limit, contributions from other regions of the integration space interfere destructively and can therefore be neglected. The critical points of all integration paths are determined by the stationary condition $\delta \mathcal{S}=0$, that is,
\begin{equation}
\frac{\partial \mathcal{S}}{\partial u_{\tau\mbox{\tiny A}}} = 
\frac{\partial \mathcal{S}}{\partial v_{\tau\mbox{\tiny A}}} = 
\frac{\partial \mathcal{S}}{\partial u_{\tau\mbox{\tiny B}}} = 
\frac{\partial \mathcal{S}}{\partial v_{\tau\mbox{\tiny B}}} = 0 ,
\end{equation}
for each time step $\tau=1,\ldots,N-1$. Taking the continuum limit $N\to\infty$, we recognize, as expected, that the set of critical points contitutes a critical path (in time) satisfying the classical equations of motion, 
\begin{equation}
\begin{array}{lll}
\displaystyle\dot{u}_{\mbox{\tiny A}}=
-\frac{i}{\hbar}\frac{\partial \tilde{H}}{\partial v_{\mbox{\tiny A}}}, &\qquad& 
\displaystyle\dot{u}_{\mbox{\tiny B}}=
-\frac{i (1+u_{\mbox{\tiny B}}v_{\mbox{\tiny B}})^2}{2j\hbar}
\frac{\partial\tilde{H}}{\partial v_{\mbox{\tiny B}}},
\\ [8pt]
\displaystyle\dot{v}_{\mbox{\tiny A}}= 
+\frac{i}{\hbar}\frac{\partial \tilde{H}}{\partial u_{\mbox{\tiny A}}},& \qquad &
\displaystyle\dot{v}_{\mbox{\tiny B}} = 
+\frac{i (1+u_{\mbox{\tiny B}}v_{\mbox{\tiny B}})^2}{2j\hbar}
\frac{\partial\tilde{H}}{\partial u_{\mbox{\tiny B}}},
\end{array}
\label{HeqM}
\end{equation}
where we define the classical Hamiltonian function $\tilde{H}=\tilde{H}(v_{\mbox{\tiny A}},v_{\mbox{\tiny B}},u_{\mbox{\tiny A}},u_{\mbox{\tiny B}})$ by
\begin{equation}
\tilde{H}(z^*,s^*,z,s) \equiv \langle z,s| \hat{H} | z,s \rangle.
\label{Hcl}
\end{equation}
To complete the specification of the trajectories, we recall that the quantum propagator depends on five parameters: $T$, $z_{\mbox{\scriptsize b}}$, $s_{\mbox{\scriptsize b}}$, $z_{\mbox{\scriptsize k}}$, and  $s_{\mbox{\scriptsize k}}$. These elements are directly involved with the boundary conditions that the critical trajectories must satisfy. In terms of the variables $[u_{\mbox{\tiny A}}(t),\,u_{\mbox{\tiny B}}(t),\,v_{\mbox{\tiny A}}(t),\,v_{\mbox{\tiny B}}(t)]$, the constraints are
\begin{equation}
\begin{array}{l}
\left.\begin{array}{ll}
u_{\mbox{\tiny A}}' \equiv u_{\mbox{\tiny A}}(0)=z_{\mbox{\scriptsize k}}, & \quad
v_{\mbox{\tiny A}}''\equiv v_{\mbox{\tiny A}}(T)=z_{\mbox{\scriptsize b}}^*, \\
u_{\mbox{\tiny B}}' \equiv u_{\mbox{\tiny B}}(0)=s_{\mbox{\scriptsize k}}, & \quad
v_{\mbox{\tiny B}}''\equiv v_{\mbox{\tiny B}}(T) =s_{\mbox{\scriptsize b}}^*,
\end{array}\right\} 
\;\mathrm{for}\;\, K_+^{\mathrm{sc}};
\\ [20pt] 
\left.\begin{array}{ll}
u_{\mbox{\tiny A}}''\equiv u_{\mbox{\tiny A}}(T)=z_{\mbox{\scriptsize k}}, & \quad
v_{\mbox{\tiny A}}' \equiv v_{\mbox{\tiny A}}(0)=z_{\mbox{\scriptsize b}}^*, \\
u_{\mbox{\tiny B}}''\equiv u_{\mbox{\tiny B}}(T)=s_{\mbox{\scriptsize k}}, & \quad
v_{\mbox{\tiny B}}' \equiv v_{\mbox{\tiny B}}(0) =s_{\mbox{\scriptsize b}}^*,
\end{array}\right\} 
\;\mathrm{for}\;\, K_-^{\mathrm{sc}}.
\end{array}
\label{bc}
\end{equation}
Notice that the trajectories involved in $K_+^{\mathrm{sc}}$ are governed by the same equations of motion of those involved in $K_-^{\mathrm{sc}}$, but they differ by their boundary conditions.

At this point, it is important to understand why the trajectories are classified as complex. Although the new variables $u_{\mbox{\tiny A}}$, $u_{\mbox{\tiny B}}$, $v_{\mbox{\tiny A}}$, and $v_{\mbox{\tiny B}}$ are complex by definition~\eqref{nv}, it is easy to see that, provided that $u_{\mbox{\tiny A}}=v^*_{\mbox{\tiny A}}$ and $u_{\mbox{\tiny B}}=v^*_{\mbox{\tiny B}}$, the classical {\em real} phase space can be recovered using the parametrization~\eqref{zs}. This reasoning would forbid us to call them complex trajectories. The crucial point is that the boundary conditions~\eqref{bc} allow for the existence of trajectories such that $u_{\mbox{\tiny A}}\neq v^*_{\mbox{\tiny A}}$ and $u_{\mbox{\tiny B}}\neq v^*_{\mbox{\tiny B}}$, since $u_{\mbox{\tiny A}}$ and $u_{\mbox{\tiny B}}$ are never constrained to be $v^*_{\mbox{\tiny A}}$ and $v^*_{\mbox{\tiny B}}$, respectively, in the same instant of time. Therefore, considering that $u_{\mbox{\tiny A}}$ and $ u_{\mbox{\tiny B}}$ are independent of $v_{\mbox{\tiny A}}$ and $v_{\mbox{\tiny B}}$, we can find {\em complex} trajectories satisfying Eqs.~\eqref{HeqM} and~\eqref{bc}. They are said to be complex in the sense that, using Eq.~\eqref{zs}, the usual classical phase-space (position, momentum) can only be recovered if we consider an extension of its variables into the complex space. Mathematically, it means that the critical paths of Eq.~\eqref{KqPI} lie in an analytically extended space. Physically, it sounds like a solution to deal with the overdetermined problem in classical phase space. Indeed, without the complex extension, it would be generally impossible to find a trajectory connecting too many boundary conditions: all phase-space coordinates are fixed at the initial and final time, as well as the evolution time. To conclude, except for the very particular case of real trajectories, where $u_{\mbox{\tiny A}}(t)=v^*_{\mbox{\tiny A}}(t)$ and $u_{\mbox{\tiny B}}(t)=v^*_{\mbox{\tiny B}}(t)$, those satisfying Eqs.~\eqref{HeqM} and~\eqref{bc} will be complex.

As we will show soon, the dynamics around a complex trajectory are crucial to defining the behavior of the semiclassical propagator. The usual way to get quantitative information from this region is through the tangent (or stability) matrix $\mathbf{M}$. This object is based on a linear approach, connecting small initial displacements with respect to the trajectory, that is, $\delta\mathbf{u}'\equiv(\delta u_{\mbox{\tiny A}}',\,\delta u_{\mbox{\tiny B}}')$ and $\delta\mathbf{v}'\equiv(\delta v_{\mbox{\tiny A}}',\,\delta v_{\mbox{\tiny B}}')$, to the final displacements $\delta\mathbf{u}''\equiv(\delta u_{\mbox{\tiny A}}'',\,\delta u_{\mbox{\tiny B}}'')$ and $\delta\mathbf{v}''\equiv(\delta v_{\mbox{\tiny A}}'',\,\delta v_{\mbox{\tiny B}}'')$. The tangent matrix is expressed by
\begin{equation}
\left(\begin{array}{c} 
\delta \mathbf{u}'' \\ \delta \mathbf{v}'' 
\end{array}\right) = 
\mathbf{M}
\left(\begin{array}{c} 
\delta \mathbf{u}' \\ \delta \mathbf{v}' 
\end{array}\right) 
= 
\left(\begin{array}{cc}
\mathbf{M}_{\mathrm{uu}} & \mathbf{M}_{\mathrm{uv}} \\
\mathbf{M}_{\mathrm{vu}} & \mathbf{M}_{\mathrm{vv}} 
\end{array}\right) 
\left(\begin{array}{c} 
\delta \mathbf{u}' \\ \delta \mathbf{v}' 
\end{array}\right) .
\label{Mtangent}
\end{equation}
Roughly speaking, we expect that unstable dynamics correspond to larger values of the $\mathbf{M}$ elements, while lower values should characterize stable trajectories. Matrix $\mathbf{M}$ can be easily calculated from a given trajectory, even numerically.

A last comment on this topic is opportune. In principle, all trajectories above specified should contribute to the semiclassical formula of $K_\pm^{\mathrm{q}} $. However, as already studied in detail, for example, in Refs.~\cite{aguiar01, ribeiro2004,ribeiro2005,ribeiro2008a,ribeiro2008b,ribeiro2011}, some of them produce nonphysical results and should be excluded from the calculation. We can understand their exclusion as follows. The success of the saddle point method depends on a deformation of the original integration paths of Eq.~\eqref{KqLI} to include the saddle points of the integrand, which give origin to the complex classical trajectories. This procedure, which is based on Cauchy's integral theorem, is generally impossible to be rigorously applied to this problem, given the number of integration variables involved. Thus, as usual, we assume that the path deformation can be performed and assign the ({\it a posteriori}) nonphysical results to the critical points that are impossible to reach by a proper path deformation.

%%%%%%%%%%%%%%%%%%%%
%%%%%%%%%%%%%%%%%%%%
\subsubsection{Semiclassical propagator formula}

Once we have identified the critical points of Eq.~\eqref{KqLI}, the last step of the saddle point method is to expand the exponent $\mathcal{S}$ around them up to second order and, thus, to perform the resulting $4N$-dimensional Gaussian integral. This procedure is the longest part of the present calculation and will not be reproduced here, given that it essentially consists of the same evaluation done in Refs.~\cite{aguiar01,ribeiro2006}. At the end of the approximation process, the semiclassical formula for the coherent-state quantum propagator~\eqref{Kq} is written as
\begin{equation}
K_\pm^{\mathrm{sc}} 
(z_{\mbox{\scriptsize b}}^*, s_{\mbox{\scriptsize b}}^*, z_{\mbox{\scriptsize k}}, s_{\mbox{\scriptsize k}}, T)
= \sum_{\mathrm{c.t.}} \;
D_\pm \, \exp{\frac{i}{\hbar} F_\pm},
\label{Ksc}
\end{equation}
where the sum indicates that all complex trajectories satisfying Eqs.~\eqref{HeqM} and~\eqref{bc} should be considered. We write the input variables of $K_\pm^{\mathrm{sc}} $ as $(z_{\mbox{\scriptsize b}}^*, s_{\mbox{\scriptsize b}}^*, z_{\mbox{\scriptsize k}}, s_{\mbox{\scriptsize k}}, T)$ to emphasize that they are the only ingredients used to determine the trajectories, as prescribed by Eq.~\eqref{bc}. The variables $z_{\mbox{\scriptsize b}}$, $s_{\mbox{\scriptsize b}}$, $z_{\mbox{\scriptsize k}}^*$, and $s_{\mbox{\scriptsize k}}^*$, on the other hand, will just enter in the normalization term~$\Lambda$, defined next, having less importance to understand $K_\pm^{\mathrm{sc}} $. We now detail the other functions appearing in Eq.~\eqref{Ksc}. We first present its exponent 
\begin{equation}
F_\pm \equiv S_\pm + G_\pm +i\hbar\Lambda,
\label{F}
\end{equation}
where the action $S_\pm$ is given by
\begin{equation}
\begin{aligned}
\frac{i}{\hbar} S_\pm (z_{\mbox{\scriptsize b}}^*, s_{\mbox{\scriptsize b}}^*, z_{\mbox{\scriptsize k}}, s_{\mbox{\scriptsize k}}, T)&= 
\pm \, \int_0^T \left[
\frac{ \dot{v}_{\mbox{\tiny A}}u_{\mbox{\tiny A}} - v_{\mbox{\tiny A}} \dot{u}_{\mbox{\tiny A}}}{2} 
+j\frac{\dot{v}_{\mbox{\tiny B}}u_{\mbox{\tiny B}} - v_{\mbox{\tiny B}} \dot{u}_{\mbox{\tiny B}}}
{1+u_{\mbox{\tiny B}}v_{\mbox{\tiny B}}}-\frac{i}{\hbar} \tilde{H} \right]\mbox{d}t \\
&\quad \,
+\frac12 \left[ u_{\mbox{\tiny A}}'\,v_{\mbox{\tiny A}}'+
u_{\mbox{\tiny A}}''\,v_{\mbox{\tiny A}}'' \right] 
+ j\ln{\big[ \big( 1+ u_{\mbox{\tiny B}}' \, v_{\mbox{\tiny B}}' \big)
\big( 1+ u_{\mbox{\tiny B}}'' \, v_{\mbox{\tiny B}}'' \big) \big]},
\end{aligned}
\label{S}
\end{equation}
the function $G_\pm$ reads 
\begin{equation}
\frac{i}{\hbar}G_\pm (z_{\mbox{\scriptsize b}}^*, s_{\mbox{\scriptsize b}}^*, z_{\mbox{\scriptsize k}}, s_{\mbox{\scriptsize k}}, T)= \pm\frac12 \int_0^{T} \left[ 
\frac{i}{\hbar}\frac{\partial^2 \tilde{H}}{\partial u_{\mbox{\tiny A}}\partial v_{\mbox{\tiny A}}}
+\frac12\left(  
\frac{\partial\dot{v}_{\mbox{\tiny B}}}{\partial v_{\mbox{\tiny B}}}- 
\frac{\partial\dot{u}_{\mbox{\tiny B}}}{\partial u_{\mbox{\tiny B}}}\right)\right]\mbox{d}t,
\label{G}
\end{equation}
and the normalization term $\Lambda$ amounts to 
\begin{equation}
\Lambda = \frac{1}{2}\left( |z_{\mbox{\scriptsize b}}|^2 +|z_{\mbox{\scriptsize k}}|^2\right) + 
j\ln{\big[ \left( 1+|s_{\mbox{\scriptsize b}}|^2 \right)\left( 1+|s_{\mbox{\scriptsize k}}|^2 \right) \big]}.
\label{Lambda}
\end{equation}
Now, we introduce the prefactor term
\begin{equation}
D_\pm (z_{\mbox{\scriptsize b}}^*, s_{\mbox{\scriptsize b}}^*, z_{\mbox{\scriptsize k}}, s_{\mbox{\scriptsize k}}, T)\equiv
\left\{  
\frac{ \big( 1+ u_{\mbox{\tiny B}}' \, v_{\mbox{\tiny B}}' \big)
\big( 1+ u_{\mbox{\tiny B}}'' \, v_{\mbox{\tiny B}}'' \big)}{2j}
\det  \left[ \frac{i}{\hbar}
\left(\begin{array}{ccc}
\displaystyle\frac{\partial^2 S_\pm}{\partial z_{\mbox{\scriptsize k}}\partial z_{\mbox{\scriptsize b}}^*} &
 \quad &
\displaystyle\frac{\partial^2 S_\pm}{\partial z_{\mbox{\scriptsize k}} \partial s_{\mbox{\scriptsize b}}^*}  \\
 [8pt]
\displaystyle\frac{\partial^2 S_\pm}{\partial s_{\mbox{\scriptsize k}} \partial z_{\mbox{\scriptsize b}}^*}  &
  \quad &
\displaystyle\frac{\partial^2 S_\pm}{\partial s_{\mbox{\scriptsize k}} \partial s_{\mbox{\scriptsize b}}^*}  
\end{array} \right) \right]
  \right\}^{1/2}.
\label{D}
\end{equation}
This expression completes our brief presentation of the semiclassical forward and backward propagators~\eqref{Ksc} in the basis $\{|z,s\rangle\}$. 

In principle, except for the second derivatives of the action seen in $D_\pm$, all functions involved in $K_\pm^{\mathrm{sc}} $ can be numerically handled. However, we can also show a way to find those derivatives in a numerical scenario. For such, we first note that the initial and final coordinates that are unconstrained by the boundary conditions~\eqref{bc}, that is,
\begin{equation}
\begin{array}{lllll}
v_{\mbox{\tiny A}}' \equiv v_{\mbox{\tiny A}}(0), & 
u_{\mbox{\tiny A}}''\equiv u_{\mbox{\tiny A}}(T), &
v_{\mbox{\tiny B}}' \equiv v_{\mbox{\tiny B}}(0), & 
\mathrm{and}\;u_{\mbox{\tiny B}}''\equiv u_{\mbox{\tiny B}}(T), &
\mathrm{for}\;\, K_+^{\mathrm{sc}},
\\
v_{\mbox{\tiny A}}''\equiv v_{\mbox{\tiny A}}(T), & 
u_{\mbox{\tiny A}}' \equiv u_{\mbox{\tiny A}}(0), &
v_{\mbox{\tiny B}}''\equiv v_{\mbox{\tiny B}}(T), & 
\mathrm{and}\;u_{\mbox{\tiny B}}' \equiv u_{\mbox{\tiny B}}(0), &
\mathrm{for}\;\, K_-^{\mathrm{sc}},
\end{array}
\nonumber
\end{equation}
can be obtained from the first derivatives of $S_\pm$. Indeed, from Eqs~\eqref{bc} and~\eqref{S}, which implicitly defines $S_+(v_{\mbox{\tiny A}}'',v_{\mbox{\tiny B}}'',u_{\mbox{\tiny A}}',u_{\mbox{\tiny B}}',T)$ and $S_-(v_{\mbox{\tiny A}}',v_{\mbox{\tiny B}}',u_{\mbox{\tiny A}}'',u_{\mbox{\tiny B}}'',T)$, we can directly find
\begin{equation}
\begin{array}{llll}
\displaystyle u_{\mbox{\tiny A}}'' = 
\frac{i}{\hbar}\frac{\partial S_+}{\partial v_{\mbox{\tiny A}}''},& \quad
\displaystyle \frac{ u_{\mbox{\tiny B}}''}{1+ u_{\mbox{\tiny B}}''v_{\mbox{\tiny B}}''  } = 
\frac{i}{2j\hbar}\frac{\partial S_+}{\partial v_{\mbox{\tiny B}}'' } ,& \quad 
\displaystyle v_{\mbox{\tiny A}}' = 
\frac{i}{\hbar}\frac{\partial S_+}{\partial u_{\mbox{\tiny A}}'},& \quad
\displaystyle \frac{ v_{\mbox{\tiny B}}'}{1+ u_{\mbox{\tiny B}}'v_{\mbox{\tiny B}}'  } = 
\frac{i}{2j\hbar}\frac{\partial S_+}{\partial u_{\mbox{\tiny B}}' } ;
\\[10pt]
\displaystyle u_{\mbox{\tiny A}}' = 
\frac{i}{\hbar}\frac{\partial S_-}{\partial v_{\mbox{\tiny A}}'},& \quad
\displaystyle \frac{ u_{\mbox{\tiny B}}'}{1+ u_{\mbox{\tiny B}}'v_{\mbox{\tiny B}}'  } = 
\frac{i}{2j\hbar}\frac{\partial S_-}{\partial v_{\mbox{\tiny B}}' } , &\quad
\displaystyle v_{\mbox{\tiny A}}'' = 
\frac{i}{\hbar}\frac{\partial S_-}{\partial u_{\mbox{\tiny A}}''},& \quad
\displaystyle \frac{ v_{\mbox{\tiny B}}''}{1+ u_{\mbox{\tiny B}}''v_{\mbox{\tiny B}}''  } = 
\frac{i}{2j\hbar}\frac{\partial S_-}{\partial u_{\mbox{\tiny B}}'' } .
\end{array}
\label{dS}
\end{equation}
Then, by differentiating these equations and rearranging the terms, we can compare the resulting expressions with the definition of the tangent matrix~\eqref{Mtangent}. This procedure is performed in App.~\ref{ap1} and, according to Eqs.~\eqref{SM+} and~\eqref{SM-}, the prefactor becomes
\begin{equation}
D_+ \equiv
\left\{  
\frac{\big( 1+ u_{\mbox{\tiny B}}'' \, v_{\mbox{\tiny B}}'' \big)}
{\big( 1+ u_{\mbox{\tiny B}}' \, v_{\mbox{\tiny B}}' \big)}
\frac{1}{\det \mathbf{M}_{\mathrm{vv}}} \right\}^{1/2}
\quad\mathrm{and}\quad
D_- \equiv
\left\{  
\frac{\big( 1+ u_{\mbox{\tiny B}}'' \, v_{\mbox{\tiny B}}'' \big)}
{\big( 1+ u_{\mbox{\tiny B}}' \, v_{\mbox{\tiny B}}' \big)}
\frac{1}{\det \mathbf{M}_{\mathrm{uu}}} \right\}^{1/2},
\label{DM}
\end{equation}
which represents an advantage for numerical applications.

Expression~\eqref{Ksc} and the saddle point method discussed above will be the crucial elements to get the semiclassical entanglement formula that we will derive in the next section.

%%%%%%%%%%%%%%%%%%%%
%%%%%%%%%%%%%%%%%%%%
\section{Semiclassical entanglement}
\label{SEsec}

Entanglement in a bipartite pure state $\hat\rho$, acting on the Hilbert space $\mathcal{H}=\mathcal{H}_z\otimes\mathcal{H}_s$, can be evaluated by the linear entropy $S^{\mathrm{q}}_{\mathrm{lin}}(\hat\rho_s) \equiv 1 - P^{\mathrm{q}}(\hat\rho_s)$ of the reduced state $\hat\rho_s\equiv\mathrm{Tr}_z \, \hat \rho$, where $P^{\mathrm{q}}(\hat\rho_s)\equiv \mathrm{Tr}_s\left\{\hat \rho_s^2\right\}$  is the purity of $\hat\rho_s$. The linear entanglement entropy of $\hat{\rho}$, defined as
\begin{equation}
E^{\mathrm{q}}(\hat\rho)\equiv S^{\mathrm{q}}_{\mathrm{lin}}(\hat\rho_s),
\label{EntQ}
\end{equation}
gives the same result if we replace $\hat\rho_s$ with $\hat\rho_z \equiv\mathrm{Tr}_s \, \hat \rho$. The function $E^{\mathrm{q}}(\hat\rho)$ is an entanglement measure because, for a separable state $\hat\rho$, we have $E^{\mathrm{q}}(\hat\rho) =0$, while an entangled state yields $0<E^{\mathrm{q}}(\hat\rho)<1$.  

We aim to study the time evolution of the entanglement when the initial state is pure and separable, explicitly given by a product of the canonical coherent state and a spin coherent state,
\begin{equation}
  \hat{\rho}_0 =  |z_0,s_0\rangle \langle z_0, s_0|.
  \label{rho0}
%\equiv
%\big[ |z_0\rangle\otimes|s_0\rangle\big]
%\big[ \langle z_0|\otimes \langle s_0|\big].
\end{equation}
Then, we evolve $\hat{\rho}_0$ until the time $T$, according to a general Hamiltonian $\hat{H}$, and calculate the time-dependent purity, now rewritten as
\begin{equation}
\begin{aligned}
P_T^{\mathrm{q}} &=\int K^{\mathrm{q}}_+ (z_1,s_1, z_0,s_0,T) \, K^{\mathrm{q}}_- (z_0,s_0, z_1,s_2,T) \\
&\quad\times \; K^{\mathrm{q}}_+ (z_2,s_2, z_0,s_0,T)\,  K^{\mathrm{q}}_- (z_0,s_0, z_2,s_1,T)~ 
\mbox{d}\nu(z_1,s_1) \, \mbox{d}\nu(z_2,s_2), 
\end{aligned}
\label{PTint}
\end{equation}
where
\begin{equation}
d\nu(z_1,s_1)\equiv \frac{2j+1}{\pi^2}\frac{\mbox{d}z_{1}^{\mbox{\tiny R}} \, \mbox{d}z_{1}^{\mbox{\tiny I}} \,
\mbox{d}s_1^{\mbox{\tiny R}} \, \mbox{d}s_{1}^{\mbox{\tiny I}} }{(1+|s_1|^2)^2}
\label{dnu}
\end{equation}
and the analog to $\mbox{d}\nu(z_2,s_2)$. Notice that, once we have deduced a semiclassical version $P_T^{\mathrm{sc}}$ for Eq.~\eqref{PTint}, we will naturally have an expression for the time-dependent semiclassical entanglement 
\begin{equation}
E_T^{\mathrm{sc}}(\hat\rho_0) \equiv 1 - P^{\mathrm{sc}}_T,
\label{ETsc}
\end{equation}
which is the main goal of the present paper.

To start the approximation, the first idea is to replace the exact propagators~\eqref{Kq} with their semiclassical counterparts~\eqref{Ksc} in Eq.~\eqref{PTint}, in close analogy to Refs.~\cite{ribeiro2010,ribeiro2012,ribeiro2019,ribeiro2021}. We substitute
\begin{equation}
\begin{aligned}
K^{\mathrm{q}}_+ (z_1,s_1, z_0,s_0,T) & \rightarrow K^{\mathrm{sc}}_1(z_1^*,s_1^*,z_0,s_0,T), \\
K^{\mathrm{q}}_- (z_0,s_0, z_1,s_2,T) & \rightarrow K^{\mathrm{sc}}_2(z_0^*,s_0^*,z_1,s_2,T), \\
K^{\mathrm{q}}_+ (z_2,s_2, z_0,s_0,T) & \rightarrow K^{\mathrm{sc}}_3(z_2^*,s_2^*,z_0,s_0,T), \\
K^{\mathrm{q}}_- (z_0,s_0, z_2,s_1,T) & \rightarrow K^{\mathrm{sc}}_4(z_0^*,s_0^*,z_2,s_1,T),
\end{aligned}
\label{KqTosc}
\end{equation}
where we emphasize that $K^{\mathrm{sc}}_1$ and $K^{\mathrm{sc}}_3$ are forward propagators, while $K^{\mathrm{sc}}_2$ and $K^{\mathrm{sc}}_4$ are backward ones. Next, we will interpret integral~\eqref{PTint} over the four complex planes, $z_1$, $z_2$, $s_1$, and $s_2$ as line integrals. This process is equivalent to that used to get Eq.~\eqref{KqLI}. Then, we rename the integration variables as
\begin{equation}
\begin{array}{llll}
z_1 \rightarrow u''_{2\mbox{\tiny A}} , & \quad
z_1^* \rightarrow v''_{1\mbox{\tiny A}}, & \quad
s_1 \rightarrow u''_{4\mbox{\tiny B}}, & \quad
s_1^* \rightarrow v''_{1\mbox{\tiny B} } , \\ 
z_2 \rightarrow u''_{4\mbox{\tiny A}}, & \quad
z_2^* \rightarrow v''_{3\mbox{\tiny A}}, & \quad
s_2 \rightarrow u''_{2\mbox{\tiny B}}, & \quad
s_2^* \rightarrow v''_{3\mbox{\tiny B}}, 
\end{array}
\end{equation}
in accordance with their role in the semiclassical propagators presented in Eq.~\eqref{KqTosc}. For example, the variable $z_1$ appears in  $K^{\mathrm{sc}}_2$ and should match the variable $u_{2\mbox{\tiny A}}$ at the final time $T$, suggesting the replacement $z_1 \rightarrow u''_{2\mbox{\tiny A}} $. Analogously, $z_1^*$ appears in $K^{\mathrm{sc}}_1$ and should be final value of variable $v_{1\mbox{\tiny A}}$, suggesting the replacement $z_1^* \rightarrow v''_{1\mbox{\tiny A}} $. This reasoning implicitly assumes that the trajectory $\big[u_{l\mbox{\tiny A}}(t),u_{l\mbox{\tiny B}}(t),v_{l\mbox{\tiny A}}(t),v_{l\mbox{\tiny B}}(t)\big]$ is the one contributing to propagator $K_l^{\mathrm{sc}}$, for $l=1,\ldots,4$, with some of their final points linked to the integration variables according to the boundary conditions~\eqref{bc}. Considering this new notation, we can approach Eq.~\eqref{PTint} to
\begin{equation}
P^{\mathrm{sc}}_T = \left(\frac{2j+1}{4\pi^2}\right)^2 
\int D_1\,D_2\,D_3\,D_4~\mbox{e}^{\Phi} ~
\frac{\mbox{d}u''_{2\mbox{\tiny A}} \, \mbox{d}v''_{1\mbox{\tiny A}} \,
\mbox{d} u''_{4\mbox{\tiny B}} \, \mbox{d} v''_{1\mbox{\tiny B}} }
{(1+ u''_{4\mbox{\tiny B}}  v''_{1\mbox{\tiny B}})^2} 
\frac{\mbox{d}u''_{4\mbox{\tiny A}} \, \mbox{d}v''_{3\mbox{\tiny A}} \,
\mbox{d} u''_{2\mbox{\tiny B}} \, \mbox{d} v''_{3\mbox{\tiny B}} }
{(1+ u''_{2\mbox{\tiny B}}  v''_{3\mbox{\tiny B}})^2} ,
\label{Psc}
\end{equation}
with $\Phi\equiv -\Lambda_{\mathrm{tot}}+\sum_{k=1}^4\left[\frac{i}{\hbar}\left(S_k+G_k\right)\right]$. Notice that
\begin{equation}
\begin{array}{ll}
S_1\equiv S_+ (v''_{1\mbox{\tiny A}},v''_{1\mbox{\tiny B}},z_0,s_0,T), &\quad
S_2\equiv S_- (z_0^*,s_0^*, u''_{2\mbox{\tiny A}},u''_{2\mbox{\tiny B}},T), \\
S_3\equiv S_+ (v''_{3\mbox{\tiny A}},v''_{3\mbox{\tiny B}},z_0,s_0,T), &\quad
S_4\equiv S_- (z_0^*,s_0^*, u''_{4\mbox{\tiny A}},u''_{4\mbox{\tiny B}},T), 
\end{array}
\end{equation}
with equivalent definitions for $G_1,\ldots,G_4$ and $D_1,\ldots,D_4$. In addition,
\begin{equation}
\Lambda_{\mathrm{tot}} \equiv 
2|z_0|^2 + 4j\ln(1+|s_0|^2) +
u''_{2\mbox{\tiny A}} v''_{1\mbox{\tiny A}} +
u''_{4\mbox{\tiny A}} v''_{3\mbox{\tiny A}} +
2j \ln\big[(1+ u''_{4\mbox{\tiny B}} v''_{1\mbox{\tiny B}} )
(1+ u''_{2\mbox{\tiny B}} v''_{3\mbox{\tiny B}} )\big].
\end{equation}
For simplicity, we omit the sum over contributing trajectories in Eq.~\eqref{Psc}, but we will return to this point opportunely.

%%%%%%%%%%%%%%%%%%%%
%%%%%%%%%%%%%%%%%%%%
\subsection{Entangled-boundary-condition trajectories}

Equation~\eqref{Psc} consists of a line integral in an 8-dimensional complex space, spanned by the vector ${\mathbf{r}}=(u''_{2\mbox{\tiny A}}, v''_{1\mbox{\tiny A}}, u''_{4\mbox{\tiny B}}, v''_{1\mbox{\tiny B}} , u''_{4\mbox{\tiny A}}, v''_{3\mbox{\tiny A}}, u''_{2\mbox{\tiny B}}, v''_{3\mbox{\tiny B}} )$. The integral provides a straightforward recipe for calculating $P_T^{\mathrm{sc}}$. First, by specifying the value of $\mathbf{r}$, we have enough information to find the complex trajectories contributing to each propagator. For example, according to Eq.~\eqref{bc}, in the case of $K_1^{\mathrm{sc}}$, the initial boudary condition is set as $u'_{1\mbox{\tiny A}}=z_0$ and $u'_{1\mbox{\tiny B}}=s_0$. By knowing the integration variables $v''_{1\mbox{\tiny A}}$ and $v''_{1\mbox{\tiny B}}$, the final constraints are also determined so that we can apply Eqs.~\eqref{HeqM} and~\eqref{bc} to obtain the trajectory. This reasoning can be extended to the other propagators, showing that, given $\mathbf{r}$, the integrand can be entirely calculated. Then, what we would still need to do is to span all prescribed values of $\mathbf{r}$, summing their contributions. This task is formally well-defined, but its implementation is, in general, practically impossible. Fortunately, similar to Eq.~\eqref{KqLI}, there are critical values of $\mathbf{r}$, corresponding to particular sets of four trajectories and their respective propagators, which enable a good approximation for~$P_T^{\mathrm{sc}}$.

We then use the saddle point method~\cite{bleistein} to solve Eq.~\eqref{Psc}. The critical points $\tilde{\mathbf{r}}$ are the solutions of the eight equations
\begin{equation}
%\textstyle
\frac{\partial \Phi}{\partial u''_{2\mbox{\tiny A}}} = 
\frac{\partial \Phi}{\partial v''_{1\mbox{\tiny A}}} =
\frac{\partial \Phi}{\partial u''_{4\mbox{\tiny B}}} =
\ldots =
\frac{\partial \Phi}{\partial v''_{3\mbox{\tiny B}}} = 0 .
\end{equation}
By performing direct calculations, with the support of Eq.~\eqref{dS}, we find the equalities
\begin{equation}
\begin{array}{llll}
\tilde{u}''_{2\mbox{\tiny A}} = \tilde{u}''_{1\mbox{\tiny A}}, & \quad 
\tilde{v}''_{1\mbox{\tiny A}} = \tilde{v}''_{2\mbox{\tiny A}}, & \quad 
\tilde{u}''_{4\mbox{\tiny B}} = \tilde{u}''_{1\mbox{\tiny B}}, & \quad 
\tilde{v}''_{1\mbox{\tiny B}} = \tilde{v}''_{4\mbox{\tiny B}}, \\ 
\tilde{u}''_{4\mbox{\tiny A}} = \tilde{u}''_{3\mbox{\tiny A}}, & \quad 
\tilde{v}''_{3\mbox{\tiny A}} = \tilde{v}''_{4\mbox{\tiny A}}, & \quad 
\tilde{u}''_{2\mbox{\tiny B}} = \tilde{u}''_{3\mbox{\tiny B}}, & \quad 
\tilde{v}''_{3\mbox{\tiny B}} = \tilde{v}''_{2\mbox{\tiny B}}.
\end{array}
\label{bc2}
\end{equation}
To get them, it is important to inform that the derivatives of functions $G_k$ are disregarded, as usually done in this kind of semiclassical approximation~\cite{aguiar01}. Remarkably, the above conditions imply that the four {\em critical} trajectories needed to calculate~\eqref{Psc} must have their final points mutually connected. Following the previous works~\cite{ribeiro2019, ribeiro2021}, we call Eq.~\eqref{bc2} as {\em entangled boundary conditions}. To understand their meaning, take, for example, the first equation $\tilde{u}''_{2\mbox{\tiny A}} = \tilde{u}''_{1\mbox{\tiny A}} $, showing that the integration variable $\tilde{u}''_{2\mbox{\tiny A}}$, which is the final point of the coordinate $\tilde{u}_{2\mbox{\tiny A}}$ of trajectory 2, should be adjusted to coincide with the last point $\tilde{u}''_{1\mbox{\tiny A}}$ of trajectory~1. On the other hand, in the third equation of~\eqref{bc2}, we see that the final coordinate $\tilde{u}''_{1\mbox{\tiny B}}$ of the same trajectory~1 should equal the integration variable $\tilde{u}''_{4\mbox{\tiny B}}$, which is a final coordinate of trajectory~4. Following this pattern, the other equalities also imply these puzzling constraints.

At first sight, we could think that no set of four trajectories could satisfy these so restrictive conditions. However, if the four trajectories were the same {\em real} trajectory starting at the center of the state $|z_0,s_0\rangle$, all necessary conditions would be satisfied. Indeed, if we set $z_0=\tilde{u}'_{l\mbox{\tiny A}}=\big[\tilde{v}'_{l\mbox{\tiny A}}\big]^*$ and $s_0=\tilde{u}'_{l\mbox{\tiny B}}=\big[\tilde{v}'_{l\mbox{\tiny B}}\big]^*$, for $l=1,\ldots,4$, so that the initial constraints, 
\begin{equation}
u'_{1\mbox{\tiny A}}=u'_{3\mbox{\tiny A}}=z_0, \quad  
u'_{1\mbox{\tiny B}}=u'_{3\mbox{\tiny B}}=s_0, \quad  
v'_{2\mbox{\tiny A}}=v'_{4\mbox{\tiny A}}=z_0^*, \quad  
v'_{2\mbox{\tiny B}}=v'_{4\mbox{\tiny B}}=s_0^*, 
\label{Ibc}
\end{equation}
are respected, then the final boundary conditions~\eqref{bc2} will be automatically satisfied, as all final points coincide. Surprisingly, besides the real trajectory, we will show later, when studying a particular system, that other sets of four complex contributing trajectories can also exist.
   
%%%%%%%%%%%%%%%%%%%%
%%%%%%%%%%%%%%%%%%%%
\subsection{Semiclassical entanglement entropy}

Having discussed the features of the set of contributing trajectories, we return to Eq.~\eqref{Psc} and expand the integrand up to second order around $\tilde{\mathbf{r}}$. We get
\begin{equation}
P^{\mathrm{sc}}_T =   
\left\{\prod_{l=1}^4
\frac{\tilde{D}_l~\mbox{e}^{\frac{i}{\hbar}\left(\tilde{S}_l+\tilde{G}_l\right) -
\frac12\left( |z_0|^2 + \tilde{u}''_{l\mbox{\tiny A}}  \tilde{v}''_{l\mbox{\tiny A}}\right)}}
{(1+|s_0|^2)^j(1+ \tilde{u}''_{l\mbox{\tiny B}}  \tilde{v}''_{l\mbox{\tiny B}})^{j}}
\right\}\, \tilde{\mathcal{I}} ,
\label{Pexp}
\end{equation}
where a tilde was inserted above the functions to indicate that they refer to the set of critical trajectories. The term $\tilde{\mathcal{I}}$ amounts to the Gaussian integral
\begin{equation}
\tilde{\mathcal{I}} \equiv \left(\frac{2j+1}{4\pi^2}\right)^2
\int \frac{\mbox{e}^{\frac12 (\mathbf{r}-\tilde{\mathbf r})~
\tilde{\mathbf{Q}}
~(\mathbf{r}-\tilde{\mathbf r})}}
{\prod_{l=1}^{4}(1+ \tilde{u}''_{l\mbox{\tiny B}}  \tilde{v}''_{l\mbox{\tiny B}})}
\mbox{d}^8\mathbf{r}=
\left[ \big( \det\tilde{\mathbf{L}}\big) \big(\det\tilde{\mathbf{Q}}\big) \right]^{-1/2},
\label{GI}
\end{equation}
where we have defined
\begin{equation}
\tilde{\mathbf{L}} \equiv \left(
\begin{array}{cccc}
\tilde{\mathbf{L}}_{1}''&0&0&0\\ 
0&\tilde{\mathbf{L}}_{2}''&0&0\\ 
0&0&\tilde{\mathbf{L}}_{3}''&0\\ 
0&0&0&\tilde{\mathbf{L}}_{4}''
\end{array}
\right),
\quad\mathrm{with}\quad
\tilde{\mathbf{L}}_l \equiv \left(
\begin{array}{cc}
1&0\\0&\frac{(1+u_{l\mbox{\tiny B}} v_{l\mbox{\tiny B}} )^2}{2j+1}  
\end{array}
\right).
\end{equation}
In addition, the matrix $\tilde{\mathbf{Q}}$ involves second derivatives of $\Phi$. After a convenient change in the order of its columns and rows, we have
\begin{equation}
\det\tilde{\mathbf{Q}}\equiv\det \frac{i}{\hbar}\left(
\begin{array}{cccc}
\tilde{\mathbf{S}}^{{1}}_{\mathrm{vv}}+ \tilde{u}''_{1\mbox{\tiny B}} \!\!{}^2 \tilde{\mathbf{A}}_1'' &  
i\hbar \, \mathbf{I}_{\text{\tiny A}}  & 
\mathbf{0} & -\tilde{\mathbf{A}}_4'' \\
i\hbar \, \mathbf{I}_{\text{\tiny A}}  & 
\tilde{\mathbf{S}}^{{2}}_{\mathrm{uu}}+ \tilde{v}''_{2\mbox{\tiny B}} \!\!{}^2 \tilde{\mathbf{A}}_2''&  
-\tilde{\mathbf{A}}_3''  & \mathbf{0} \\
\mathbf{0}  &  -\tilde{\mathbf{A}}_2''  & 
\tilde{\mathbf{S}}^{3}_{\mathrm{vv}} + \tilde{u}''_{3\mbox{\tiny B}} \!\!{}^2 \tilde{\mathbf{A}}_3''  & 
i\hbar \, \mathbf{I}_{\text{\tiny A}}   \\
-\tilde{\mathbf{A}}_1''  &  \mathbf{0}  & 
i\hbar \, \mathbf{I}_{\text{\tiny A}}   &  
\tilde{\mathbf{S}}^{4}_{\mathrm{uu}} +\tilde{v}''_{4\mbox{\tiny B}} \!\!{}^2\tilde{\mathbf{A}}_4'' \\
\end{array}
\right),
\label{matriz_q}
\end{equation}
where
\begin{equation}
\mathbf{I}_{\mbox{\tiny A}}\equiv
\left(\begin{array}{cc}1& 0 \\ 0 &0
\end{array}\right),\quad
\tilde{\mathbf{A}}_l \equiv \left(
\begin{array}{cc} 0&0\\
0&\frac{-2ij\hbar }{(1+\tilde{v}_{l\mbox{\tiny B}} \tilde{u}_{l\mbox{\tiny B}})^2} 
\end{array}\right),\quad\mathrm{and}\quad
\tilde{\mathbf{S}}^{1}_{\mathrm{vv}} \equiv \left(
\begin{array}{cc}
\frac{\partial^{2}\tilde S_1}{\partial v''_{1\mbox{\tiny A}} \partial v''_{1\mbox{\tiny A}}} &
\frac{\partial^{2}\tilde S_1}{\partial v''_{1\mbox{\tiny A}} \partial v''_{1\mbox{\tiny B}}} \\
\frac{\partial^{2}\tilde S_1}{\partial v''_{1\mbox{\tiny B}} \partial v''_{1\mbox{\tiny A}}} &
\frac{\partial^{2}\tilde S_1}{\partial v''_{1\mbox{\tiny B}} \partial v''_{1\mbox{\tiny B}}} 
\end{array}\right).
\end{equation}
Matrices $\mathbf{S}^{2}_{\mathrm{uu}}$, $\mathbf{S}^{3}_{\mathrm{vv}}$, and $\mathbf{S}^{4}_{\mathrm{uu}}$ are defined in analogy to $\mathbf{S}^{1}_{\mathrm{vv}}$. Now, we will insert Eq.~\eqref{GI} in~\eqref{Pexp} and use Eq.~\eqref{DM} to express the terms $D_l$. Then, we replace the second derivatives of the action with the elements of the tangent matrix according to Eqs.~\eqref{SM+} and~\eqref{SM-} and approach $2j\approx2j+1$. 

Finally, considering all these steps, the semiclassical entanglement entropy~\eqref{ETsc} becomes
\begin{equation}
E^{\mathrm{sc}}_T(\hat{\rho}_0) =  1 - 
\sum_{\mathrm{sets}}
\mathcal{A}\, 
\mbox{e}^{\frac{i}{\hbar}\left(F_1-F_2+F_3-F4\right)},
\label{Efinal}
\end{equation}
where
\begin{equation}
\frac{i}{\hbar} F_l  \equiv 
\int_0^T \left[
  \frac12 \left( \alpha_{l\mbox{\tiny A}} +
  \frac{2j \alpha_{l\mbox{\tiny B}}}{1+u_{l\mbox{\tiny B}}v_{l\mbox{\tiny B}}} \right)
  + \frac14 (\beta_{l\mbox{\tiny A}}  + \beta_{l\mbox{\tiny B}})
-\frac{i}{\hbar} \tilde{H} \right] \mbox{d}t , 
\nonumber
\end{equation}
with
\begin{equation}
  \alpha_{l\mbox{\tiny A}} \equiv
  \dot{v}_{l\mbox{\tiny A}}u_{l\mbox{\tiny A}} - v_{l\mbox{\tiny A}} \dot{u}_{l\mbox{\tiny A}}
  \quad\mathrm{and}\quad
  \beta_{l\mbox{\tiny A}} \equiv
  \frac{\partial\dot{v}_{l\mbox{\tiny A}}}{\partial v_{l\mbox{\tiny A}}}- 
  \frac{\partial\dot{u}_{l\mbox{\tiny A}}}{\partial u_{l\mbox{\tiny A}}},
\nonumber
\end{equation}
and the equivalent for part $B$. In addition,
\begin{equation}
\mathcal{A}  \equiv \sqrt{\frac{1}{\det\mathbf{R} }}~
\prod_{l=1}^4\left[
\sqrt{\frac{(1+ {u}''_{l\mbox{\tiny B}}  {v}''_{l\mbox{\tiny B}})}
{(1+ {u}'_{l\mbox{\tiny B}}  {v}'_{l\mbox{\tiny B}})}}~
\left( \frac{1+ {u}'_{l\mbox{\tiny B}}  {v}'_{l\mbox{\tiny B}}}{1+|s_0|^2} \right)^j
\mbox{e}^{-\frac12\left( |z_0|^2 - {u}'_{l\mbox{\tiny A}}  {v}'_{l\mbox{\tiny A}}\right)}\right] ,
\nonumber
\end{equation}
where matrix $\mathbf{R}$ is defined as
\begin{equation} 
\mathbf{R}  \equiv \left(
\begin{array}{cccc}
\mathbf{M}_{\mathrm{uv}}^1 & 
-\mathbf{I}_{\mbox{\tiny A}} \mathbf{M}_{\mathrm{uu}}^2  &
0 &
-\mathbf{I}_{\mbox{\tiny B}} \mathbf{M}_{\mathrm{uu}}^4 \\ 
-\mathbf{I}_{\mbox{\tiny A}} \mathbf{M}_{\mathrm{vv}}^1 & 
\mathbf{M}_{\mathrm{vu}}^2 & 
-\mathbf{I}_{\mbox{\tiny B}} \mathbf{M}_{\mathrm{vv}}^3 &
0\\ 
0&
-\mathbf{I}_{\mbox{\tiny B}} \mathbf{M}_{\mathrm{uu}}^2 &
\mathbf{M}_{\mathrm{uv}}^3 &
-\mathbf{I}_{\mbox{\tiny A}} \mathbf{M}_{\mathrm{uu}}^4 \\ 
-\mathbf{I}_{\mbox{\tiny B}} \mathbf{M}_{\mathrm{vv}}^1 &
0&
-\mathbf{I}_{\mbox{\tiny A}} \mathbf{M}_{\mathrm{vv}}^3 & 
\mathbf{M}_{\mathrm{vu}}^4
\end{array}
\right), \qquad \mathrm{with}~~
\mathbf{I}_{\mbox{\tiny B}}\equiv
\left(\begin{array}{cc}0& 0 \\ 0 &1
\end{array}\right).
\label{matR}
\end{equation}

Equation~\eqref{Efinal} is our final expression for the semiclassical entanglement entropy. It should be clear that it depends on sets of four complex trajectories governed by Eq.~\eqref{HeqM}, satisfying the initial constraints given by~\eqref{Ibc} and the final entangled boundary conditions~\eqref{bc2}. To indicate that more than one trajectory set may contribute, we included a summation in expression~\eqref{Efinal}. This inclusion naturally corrects the deliberate exclusion of the summation in Eq.~\eqref{Psc}. Also, to simplify the notation, we remove the tilde from the functions involved in $E^{\mathrm{sc}}_T(\hat{\rho}_0)$. At last, notice that, if we only take into account the set of four identical real trajectories, then Eq.~\eqref{Efinal} will reduce to $(1+ {u}''_{\mbox{\tiny B}r}  {v}''_{\mbox{\tiny B}r})^2/(1+|s_0|^2)^2/ \sqrt{\det\mathbf{R}_{r} }$, where we include the index $r$ to indicate {\em real trajectory}. It shows that, without complex trajectories, the entanglement dynamics will be essentially governed by the tangent matrix of the central classical trajectory and will somehow manifest how unstable/stable this trajectory is. As shown in the example below, including complex trajectories implies that richer entanglement behaviors can be semiclassically described.

%%%%%%%%%%%%%%%%%%%%
%%%%%%%%%%%%%%%%%%%%
\section{Example}
\label{Esec}

To help us understand how to get the set of complex trajectories and calculate $E_T^{\mathrm{sc}}$, we will consider a simple interacting system consisting of a Hamilton operator given by
\begin{equation}
\hat{H}_{\mbox{\footnotesize ex}} =  (\hbar/j) \lambda \, \hat{a}^\dagger \hat{a} \, \hat{J}_{\mathrm z} .
\label{Hex}
\end{equation}
We then calculate the classical Hamiltonian by performing the average prescribed by Eq.~\eqref{Hcl}, 
\begin{equation}
%  \langle z,s|\hat{H}|z,s\rangle =- \hbar
%\lambda \, z^* z \left( \frac{1-s^* s}{1+s^* s} \right) 
%\quad \Longrightarrow \quad
\tilde{H}_{\mbox{\footnotesize ex}} 
(v_{\mbox{\tiny A}},v_{\mbox{\tiny B}},u_{\mbox{\tiny A}},u_{\mbox{\tiny B}}) = - \hbar
\lambda \, v_{\mbox{\tiny A}}u_{\mbox{\tiny A}}  
\left(\frac{1-v_{\mbox{\tiny B}}u_{\mbox{\tiny B}}}{1+v_{\mbox{\tiny B}}u_{\mbox{\tiny B}}} \right).
\label{Htildeex}
\end{equation}
Next step is to solve Eq.~\eqref{HeqM}, which, in this example, becomes
\begin{equation}
\begin{array}{ll}
\displaystyle \dot{u}_{\mbox{\tiny A}} = + i \lambda   
\left( \frac{1-v_{\mbox{\tiny B}}u_{\mbox{\tiny B}}}
{1+v_{\mbox{\tiny B}}u_{\mbox{\tiny B}}} \right) u_{\mbox{\tiny A}},  &\quad
\displaystyle \dot{u}_{\mbox{\tiny B}} = - i (\lambda/j) v_{\mbox{\tiny A}}u_{\mbox{\tiny A}} 
u_{\mbox{\tiny B}}, \\ [8pt]
\displaystyle \dot{v}_{\mbox{\tiny A}} = - i \lambda   
\left( \frac{1-v_{\mbox{\tiny B}}u_{\mbox{\tiny B}}}
{1+v_{\mbox{\tiny B}}u_{\mbox{\tiny B}}} \right) v_{\mbox{\tiny A}},  &\quad
\displaystyle \dot{v}_{\mbox{\tiny B}} = + i (\lambda/j) v_{\mbox{\tiny A}}u_{\mbox{\tiny A}} 
v_{\mbox{\tiny B}}.
\end{array}
\label{eqmEx}
\end{equation}
Using these equations, we can easily take the time derivative of the products $v_{\mbox{\tiny A}}u_{\mbox{\tiny A}}$ and $v_{\mbox{\tiny B}}u_{\mbox{\tiny B}}$ to check that they are time-independent. From this result, we easily conclude that the complex trajectories satisfying Eq.~\eqref{eqmEx} can be expressed by
\begin{equation}
u_{\mbox{\tiny A}}(t) \propto \mbox{e}^{+i\omega_{\mbox{\tiny B}} t}, \quad 
v_{\mbox{\tiny A}}(t) \propto \mbox{e}^{-i\omega_{\mbox{\tiny B}}  t}, \quad 
u_{\mbox{\tiny B}}(t) \propto \mbox{e}^{-i\omega_{\mbox{\tiny A}}  t}, \quad 
v_{\mbox{\tiny B}}(t) \propto \mbox{e}^{+i\omega_{\mbox{\tiny A}}  t},
\label{HeqMex}
\end{equation}
where $\omega_{\mbox{\tiny A}}\equiv (\lambda/j) v_{\mbox{\tiny A}}u_{\mbox{\tiny A}}$ and $\omega_{\mbox{\tiny B}}\equiv \lambda   \left(1-v_{\mbox{\tiny B}}u_{\mbox{\tiny B}}\right)/\left(1+v_{\mbox{\tiny B}}u_{\mbox{\tiny B}} \right) $. Writing the trajectories in terms of the initial coordinates is an easy task:
\begin{equation}
u_{\mbox{\tiny A}}(t) = u_{\mbox{\tiny A}}' \mbox{e}^{+i\omega_{\mbox{\tiny B}}' t}, \quad 
v_{\mbox{\tiny A}}(t) = v_{\mbox{\tiny A}}' \mbox{e}^{-i\omega_{\mbox{\tiny B}}'  t}, \quad 
u_{\mbox{\tiny B}}(t) = u_{\mbox{\tiny B}}' \mbox{e}^{-i\omega_{\mbox{\tiny A}}'  t}, \quad 
v_{\mbox{\tiny B}}(t) = v_{\mbox{\tiny B}}' \mbox{e}^{+i\omega_{\mbox{\tiny A}}'  t},
\label{HeqMexl}
\end{equation}
where $\omega_{\mbox{\tiny A}}' = (\lambda/j) v_{\mbox{\tiny A}}'u_{\mbox{\tiny A}}'$ and $\omega_{\mbox{\tiny B}}' = \lambda   \left(1-v_{\mbox{\tiny B}}'u_{\mbox{\tiny B}}'\right)/\left(1+v_{\mbox{\tiny B}}'u_{\mbox{\tiny B}}' \right) $. However, these equations cannot be directly used to find the sets of contributing trajectories to Eq.~\eqref{Efinal}, as their boundary conditions involve the final time $T$. The advantage of Eq.~\eqref{HeqMexl}, however, is that they can be differentiated to find the tangent matrix~\eqref{Mtangent}:
\begin{equation}
\left(\begin{array}{c}
\delta u_{\mbox{\tiny A}}''\\\delta u_{\mbox{\tiny B}}''\\
\delta v_{\mbox{\tiny A}}''\\\delta v_{\mbox{\tiny B}}''
\end{array}\right)=
\left(\begin{array}{cccc}
\frac{u_{\mbox{\tiny A}}''}{u_{\mbox{\tiny A}}'} %\mbox{e}^{+i\omega_{\mbox{\tiny B}}' T} 
& 
\frac{-2i\lambda T u_{\mbox{\tiny A}}'' v_{\mbox{\tiny B}}'}{(1+u_{\mbox{\tiny B}}' v_{\mbox{\tiny B}}')^2} &
0&
\frac{-2i\lambda T u_{\mbox{\tiny A}}'' u_{\mbox{\tiny B}}'}{(1+u_{\mbox{\tiny B}}' v_{\mbox{\tiny B}}')^2} 
\\
\frac{-i \lambda T u_{\mbox{\tiny B}}'' v_{\mbox{\tiny A}}'}{j} &
\frac{u_{\mbox{\tiny B}}''}{u_{\mbox{\tiny B}}'} % \mbox{e}^{-i\omega_{\mbox{\tiny A}}' T} 
& 
\frac{-i \lambda  T u_{\mbox{\tiny B}}'' u_{\mbox{\tiny A}}'}{j} &
0
\\
0&
\frac{+2i\lambda T v_{\mbox{\tiny A}}'' v_{\mbox{\tiny B}}'}{(1+u_{\mbox{\tiny B}}' v_{\mbox{\tiny B}}')^2} &
\frac{v_{\mbox{\tiny A}}''}{v_{\mbox{\tiny A}}'}% \mbox{e}^{-i\omega_{\mbox{\tiny B}}' T} 
& 
\frac{+2i\lambda T v_{\mbox{\tiny A}}'' u_{\mbox{\tiny B}}'}{(1+u_{\mbox{\tiny B}}' v_{\mbox{\tiny B}}')^2} 
\\
\frac{+i \lambda T v_{\mbox{\tiny B}}'' v_{\mbox{\tiny A}}'}{j} &
0&
\frac{+i \lambda T v_{\mbox{\tiny B}}'' u_{\mbox{\tiny A}}'}{j} &
\frac{v_{\mbox{\tiny B}}''}{v_{\mbox{\tiny B}}'}% \mbox{e}^{+i\omega_{\mbox{\tiny A}}' T}
\end{array}\right)
\left(\begin{array}{c}
\delta u_{\mbox{\tiny A}}'\\\delta u_{\mbox{\tiny B}}'\\
\delta v_{\mbox{\tiny A}}'\\\delta v_{\mbox{\tiny B}}'
\end{array}\right) .
\label{TangEx}
\end{equation}

To find the sets of contributing trajectories, we apply the initial conditions~\eqref{Ibc} to trajectories~\eqref{HeqMexl} and impose the final entanglement boundary conditions~\eqref{bc2}, getting
\begin{equation}
\begin{array}{lll}
u''_{2\mbox{\tiny A}} = u''_{1\mbox{\tiny A}} 
&\Longrightarrow&
u'_{2\mbox{\tiny A}} \exp{+i \lambda
\left( \frac{1- s_0^* u'_{2 \mbox{\tiny B}}}{1+ s_0^* u'_{2\mbox{\tiny B}}} \right)T} = 
z_0 \exp{+i \lambda
\left( \frac{1- v'_{1 \mbox{\tiny B}} s_0}{1+ v'_{1\mbox{\tiny B}} s_0} \right)T} ,
\\
v''_{1\mbox{\tiny A}} = v''_{2\mbox{\tiny A}} &\Longrightarrow&
v'_{1\mbox{\tiny A}} \exp{-i \lambda
\left( \frac{1- v'_{1 \mbox{\tiny B}} s_0}{1+ v'_{1\mbox{\tiny B}} s_0} \right)T} = 
z_0^* \exp{-i \lambda
\left( \frac{1- s_0^* u'_{2 \mbox{\tiny B}}}{1+ s_0^* u'_{2\mbox{\tiny B}}} \right)T} ,
\\ 
u''_{4\mbox{\tiny A}} = u''_{3\mbox{\tiny A}} 
&\Longrightarrow&
u'_{4\mbox{\tiny A}} \exp{+i \lambda
\left( \frac{1- s_0^* u'_{4 \mbox{\tiny B}}}{1+ s_0^* u'_{4\mbox{\tiny B}}} \right)T} = 
z_0 \exp{+i \lambda
\left( \frac{1- v'_{3 \mbox{\tiny B}} s_0}{1+ v'_{3\mbox{\tiny B}} s_0} \right)T} ,
\\
v''_{3\mbox{\tiny A}} = v''_{4\mbox{\tiny A}} 
&\Longrightarrow&
v'_{3\mbox{\tiny A}} \exp{-i \lambda
\left( \frac{1- v'_{3 \mbox{\tiny B}} s_0}{1+ v'_{3\mbox{\tiny B}} s_0} \right)T} = 
z_0^* \exp{-i \lambda
\left( \frac{1- s_0^* u'_{4 \mbox{\tiny B}}}{1+ s_0^* u'_{4\mbox{\tiny B}}} \right)T} ,
\\ 
u''_{4\mbox{\tiny B}} = u''_{1\mbox{\tiny B}} 
&\Longrightarrow&
u'_{4\mbox{\tiny B}} \exp{-i \lambda z_0^* u'_{4\mbox{\tiny A}} T/j} = 
s_0 \exp{-i \lambda v'_{1\mbox{\tiny A}} z_0 T/j} ,
\\
v''_{1\mbox{\tiny B}} = v''_{4\mbox{\tiny B}} 
&\Longrightarrow&
v'_{1\mbox{\tiny B}} \exp{+i \lambda v'_{1\mbox{\tiny A}} z_0 T/j} = 
s_0^*  \exp{+i \lambda z_0^* u'_{4\mbox{\tiny A}} z_0 T/j} ,
\\
u''_{2\mbox{\tiny B}} = u''_{3\mbox{\tiny B}} 
&\Longrightarrow&
u'_{2\mbox{\tiny B}} \exp{-i \lambda z_0^* u'_{2\mbox{\tiny A}} T/j} = 
s_0 \exp{-i \lambda v'_{3\mbox{\tiny A}} z_0 T/j} ,
\\
v''_{3\mbox{\tiny B}} = v''_{2\mbox{\tiny B}} 
&\Longrightarrow&
v'_{3\mbox{\tiny B}} \exp{+i \lambda v'_{3\mbox{\tiny A}} z_0 T/j} = 
s_0^*  \exp{+i \lambda z_0^* u'_{2\mbox{\tiny A}} z_0 T/j} .
\end{array}
\label{bc2ex}
\end{equation}
Notice that these eight equalities allow us to seek the eight initial coordinates needed to define a set of four contributing trajectories. By renaming the variables as
\begin{equation}
\begin{array}{llll}
v'_{1\mbox{\tiny A}} \equiv z_0^* \alpha_1, & \quad
v'_{1\mbox{\tiny B}} \equiv s_0^* \beta_1, & \quad
u'_{2\mbox{\tiny A}} \equiv z_0 \alpha_2, & \quad
v'_{2\mbox{\tiny B}} \equiv s_0 \beta_2, 
\\
v'_{3\mbox{\tiny A}} \equiv z_0^* \alpha_3, & \quad
v'_{3\mbox{\tiny B}} \equiv s_0^* \beta_3, & \quad
u'_{4\mbox{\tiny A}} \equiv z_0 \alpha_4, & \quad
v'_{4\mbox{\tiny B}} \equiv s_0 \beta_4, 
\end{array}
\end{equation}
and manipulating Eq.~\eqref{bc2ex}, we get an equation for $\alpha_1$:
\begin{equation}
f(\alpha_1) \equiv f_s\big[ f_z(\alpha_1)\big] -\alpha_1 = 0,
\label{TranscEq}
\end{equation}
where %$f_s(x) \equiv \mbox{e}^{g_s(x)}$ and $f_z(x) \equiv \mbox{e}^{g_z(x)}$, with
\begin{equation}
f_s(x) \equiv \exp{\frac{-2i\lambda |s_0|^2 (x^2-1)T}{(1+|s_0|^2 x)(x+|s_0|^2)}}
\quad\mathrm{and}\quad
f_z(x) \equiv \exp{\frac{-i\lambda |z_0|^2 (x^2-1)T}{jx}}.
\end{equation}
Once we have a solution for Eq.~\eqref{TranscEq}, the other unknown variables are calculated through
\begin{equation}
\alpha_1 = \alpha_2 = \frac{1}{\alpha_3} = \frac{1}{\alpha_4} , 
\quad\mathrm{and}\quad
f_z (\alpha_1) = \beta_1 = \frac{1}{\beta_2} = \frac{1}{\beta_3} = \beta_4 .
\end{equation}
To understand the transcendental equation~\eqref{TranscEq}, notice that, for $T=0$, only $\alpha_1 = 1$ is a solution. This root implies that $\alpha_l=\beta_l=1$ (for $l=1,\ldots,4$), generating the set of four identical {\em real} trajectories. Most importantly, $\alpha_1=1$ is a root of Eq.~\eqref{TranscEq} whatever is the value of $T$, as expected, since we already discussed that the (central) real trajectory always contributes to $E_T^{\mathrm{sc}}$. As we will show numerically next, as $T$ increases, many other relevant solutions arise. 

We point out that, for this simple system, the problem of describing the semiclassical entanglement dynamics essentially becomes a task of finding the roots of the transcendental equation~\eqref{TranscEq}. From the values $\alpha_1$, we simply evaluate the four trajectories to insert in Eq.~\eqref{Efinal}, getting
\begin{equation}
E^{\mathrm{sc}}_T(\hat{\rho}_0) =  1 - 
\sum_{\alpha_1} 
\mathcal{A} ~ 
\exp{ -\frac{i\,j}{\lambda T }~g_z(\alpha_1)~g_s\big[f_z(\alpha_1)\big] },
\label{EfinalEx}
\end{equation}
where
\begin{equation}
g_s(x) \equiv \frac{-2i\lambda |s_0|^2 (x^2-1)T}{(1+|s_0|^2 x)(x+|s_0|^2)}, \qquad
g_z(x) \equiv \frac{-i\lambda |z_0|^2 (x^2-1)T}{jx},
\end{equation}
and
\begin{equation}
\mathcal{A} \equiv  
\left[ \frac{(1+f_z(\alpha_1)|s_0|^2)(f_z(\alpha_1)+|s_0|^2) }{f_z(\alpha_1)(1+|s_0|^2)^2}\right]^{2j}
\frac{\exp{|z_0|^2 \left[ 2  - (\alpha_1^2 + 1)/\alpha_1\right] }}
{\sqrt{\det\mathbf{R}}} .
\end{equation}
Here, the only term that still needs to be calculated is the matrix $\mathbf{R}$, defined by Eq.~\eqref{matR}, but their ingredients are already explicitly shown in Eq.~\eqref{TangEx}.

Before presenting our numerical results, we will briefly review the entirely quantum entanglement entropy. Straightforward calculations applied to Eq.~\eqref{EntQ}, using the initial state~\eqref{rho0} and the Hamiltonian~\eqref{Hex} yield
\begin{equation}
  E_T^{\mathrm{q}}(\hat\rho_0) =
  \mathcal{N}_{z_0}^4   \, \mathcal{N}_{s_0}^4
  \sum_{n_z,m_z=0}^\infty \!\!
  \mathcal{A}_{n_z}^{(z_0)} \, \mathcal{A}_{m_z}^{(z_0)}
  \sum_{n_s,m_s=0}^{2j}\!\!
  \mathcal{B}_{n_s}^{(s_0)} \, \mathcal{B}_{m_s}^{(s_0)}\,
  \mbox{e}^{-i(\lambda T/j)(n_z-m_z)(n_s-m_s)},
  \label{ETqEx}
\end{equation}
where
\begin{equation}
  \mathcal{A}_{n}^{(z_0)} \equiv \frac{|z_0|^{2n}}{n!}
  \quad\mathrm{and}\quad
  \mathcal{B}_{m}^{(s_0)} \equiv \frac{(2j)!|s_0|^{2m}}{(2j-m)!m!}.
\end{equation}
For our purposes, it is convenient to realize that $E^{\mathrm{q}}(\hat\rho_0)$ is periodic with period $T_\mathrm{p} = 2\pi j/\lambda$. Therefore, we define the dimensionless time
\begin{equation}
  \tau = \frac{T}{T_\mathrm{p}},
\end{equation}
so that the whole entanglement dynamics is contained in the interval $0\le \tau <1$. In the next section, all the discussion will be done in terms of $\tau$, instead of $T$, and we will investigate the whole period, $0\le \tau <1$. 

%%%%%%%%%%%%%%%%%%%%
%%%%%%%%%%%%%%%%%%%%
\subsection{Numerical calculation}

By inspecting Eq.~\eqref{TranscEq}, we first notice that
\begin{equation}
  \mathrm{if}\;
  f(\bar\alpha_1) = 0,\;
  \mathrm{then}\;
  f(\bar\alpha_1^*)=f(1/\bar\alpha_1)=f(1/\bar\alpha_1^*)=0.
  \label{sym}
\end{equation}
Therefore, in the $\alpha_1$ complex plane, we just need to look for roots of $f(\alpha_1)$ inside the unit circle and restricted to $\alpha_1^{\text{\tiny I}} \ge 0$. The solutions of Eq.~\eqref{TranscEq} outside this region can be straightforwardly found by applying the operations~\eqref{sym}.

\begin{figure}
\centerline{
	\includegraphics[width=6cm]{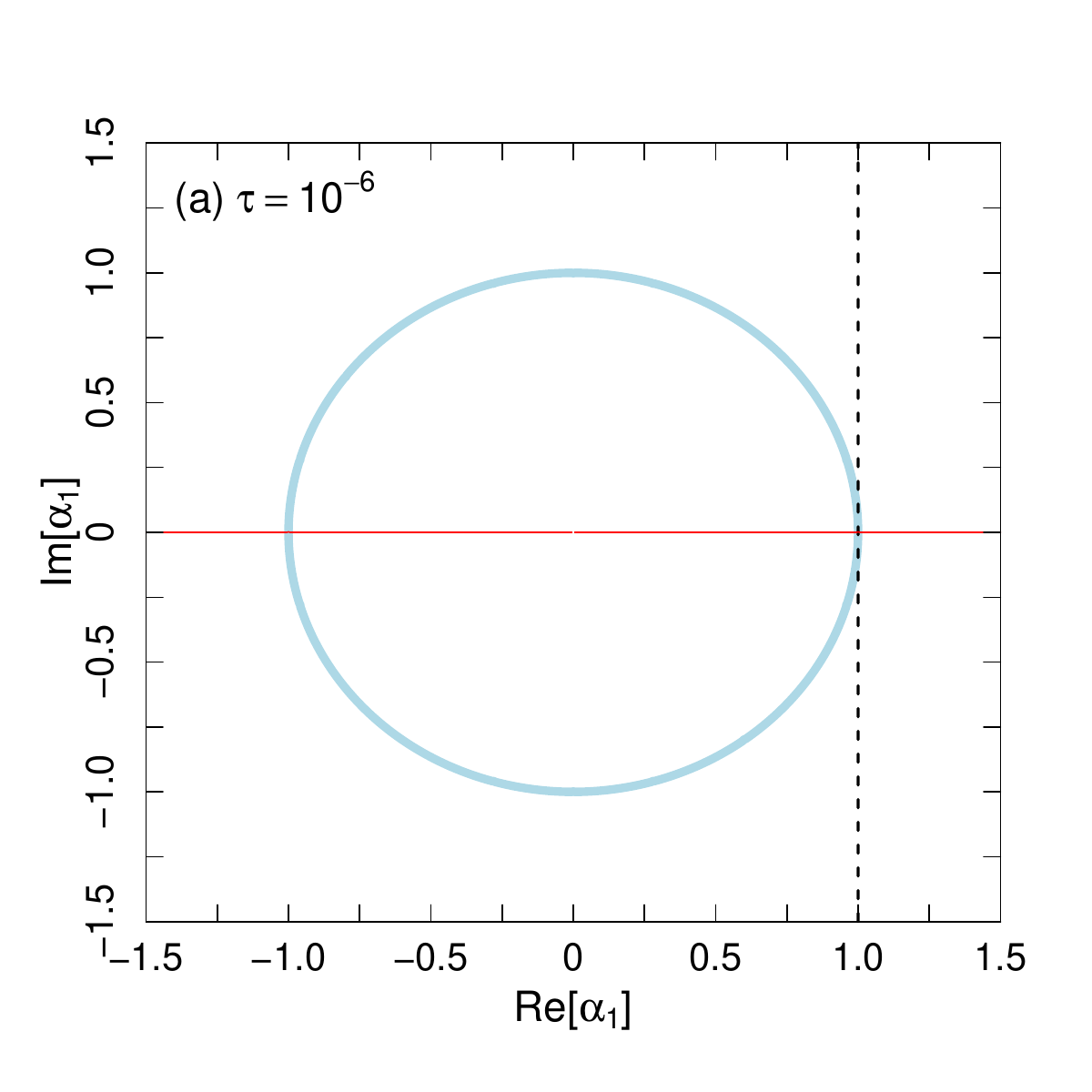} \hspace{-0.5cm}
	\includegraphics[width=6cm]{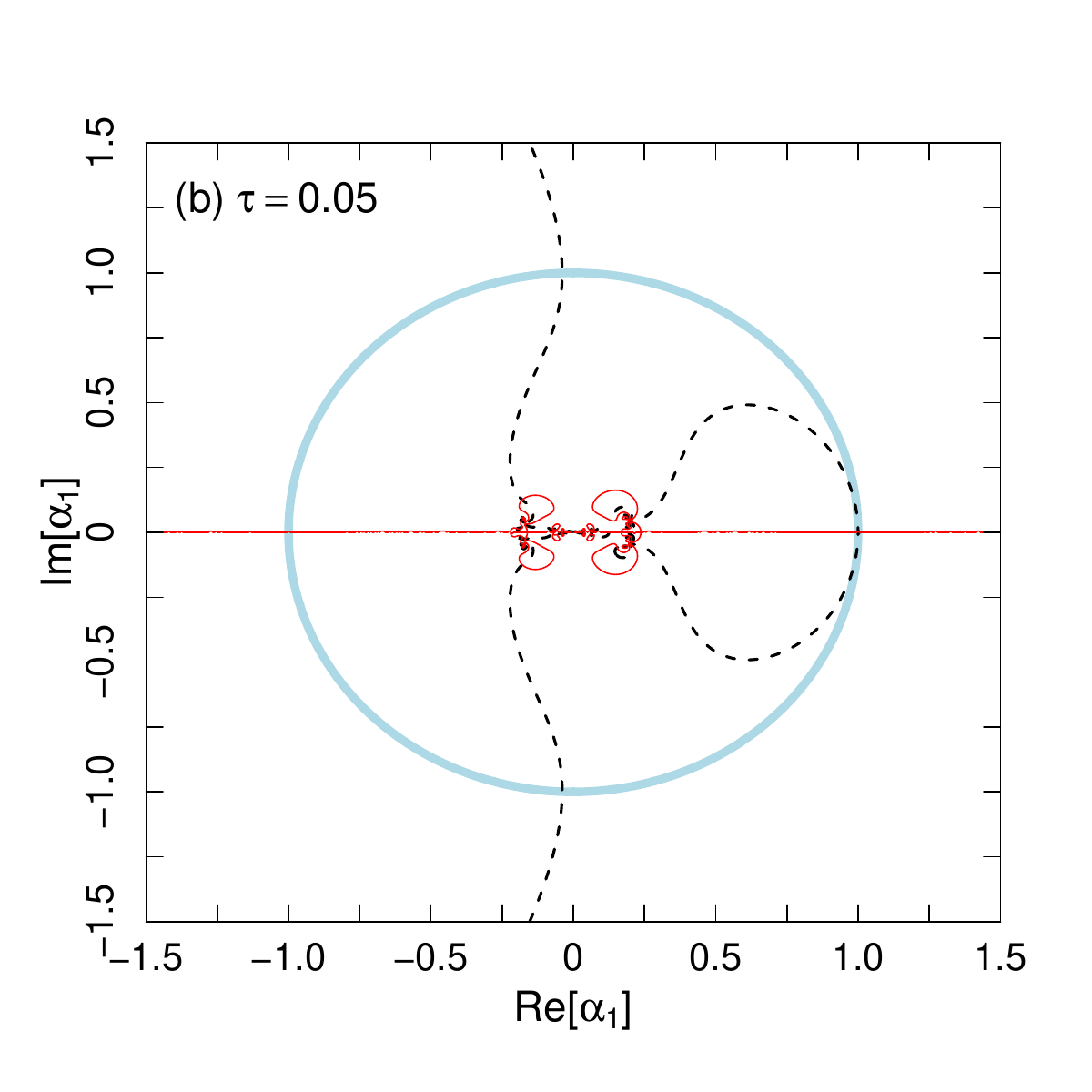} \hspace{-0.5cm}
	\includegraphics[width=6cm]{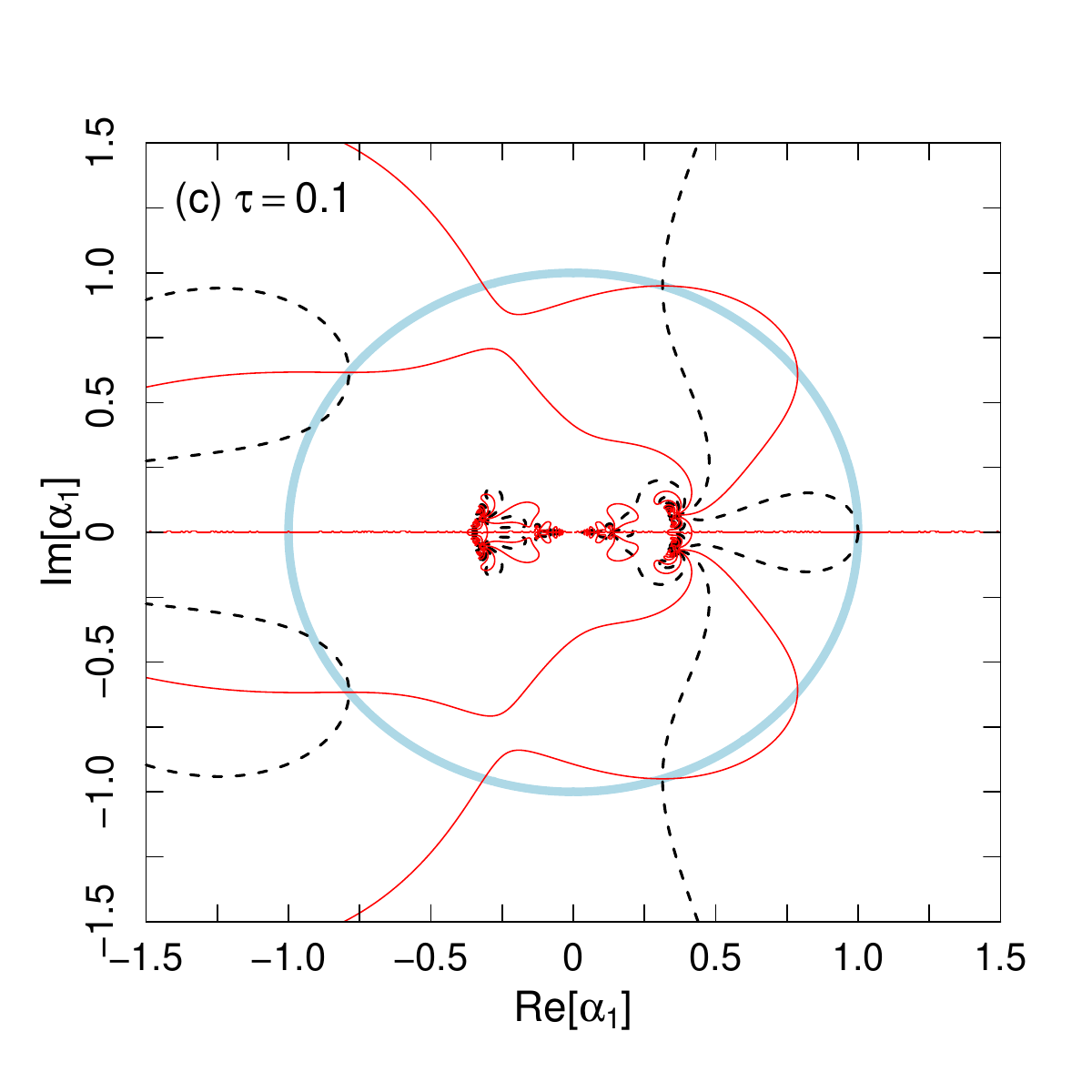}}
\centerline{
	\includegraphics[width=6cm]{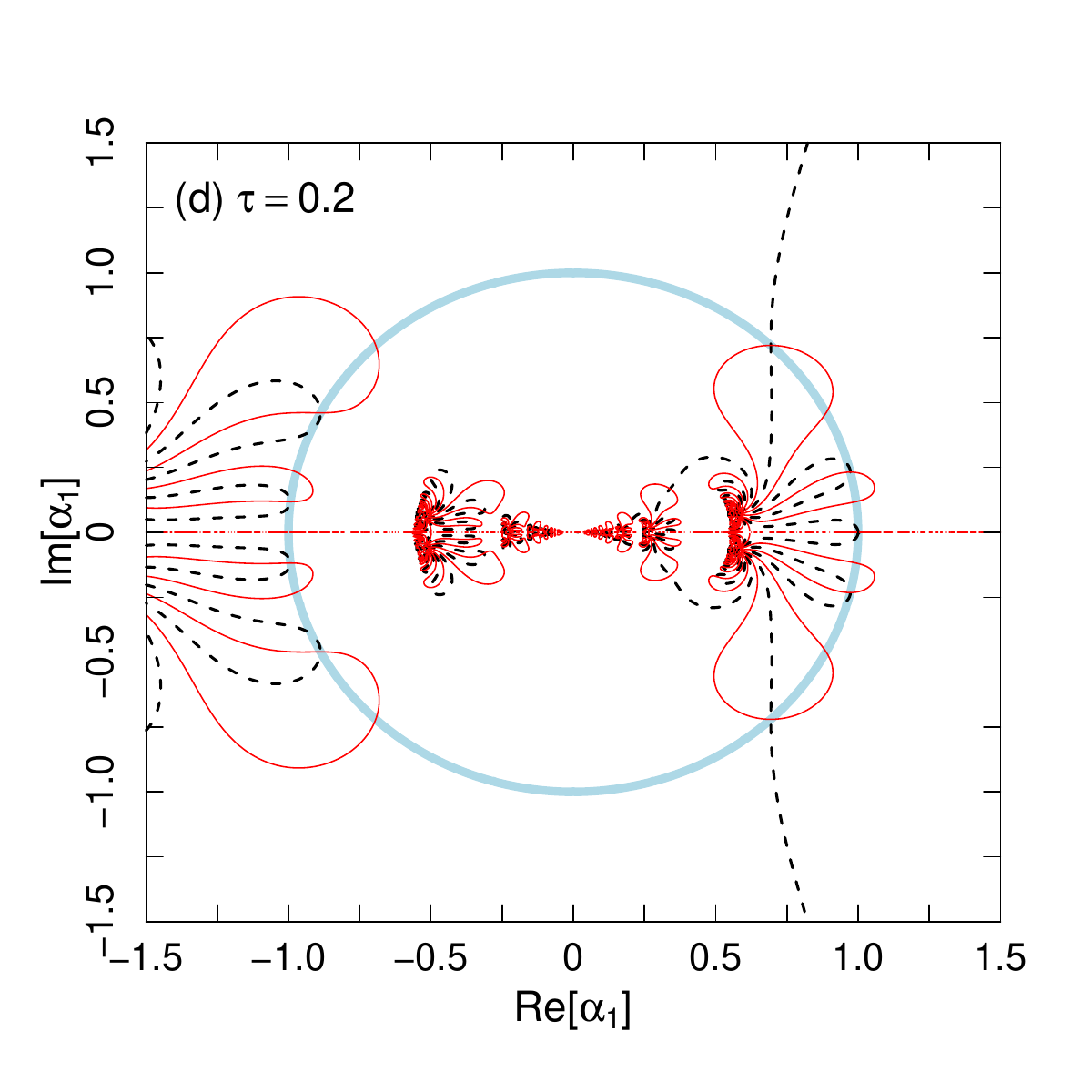} \hspace{-0.5cm}
	\includegraphics[width=6cm]{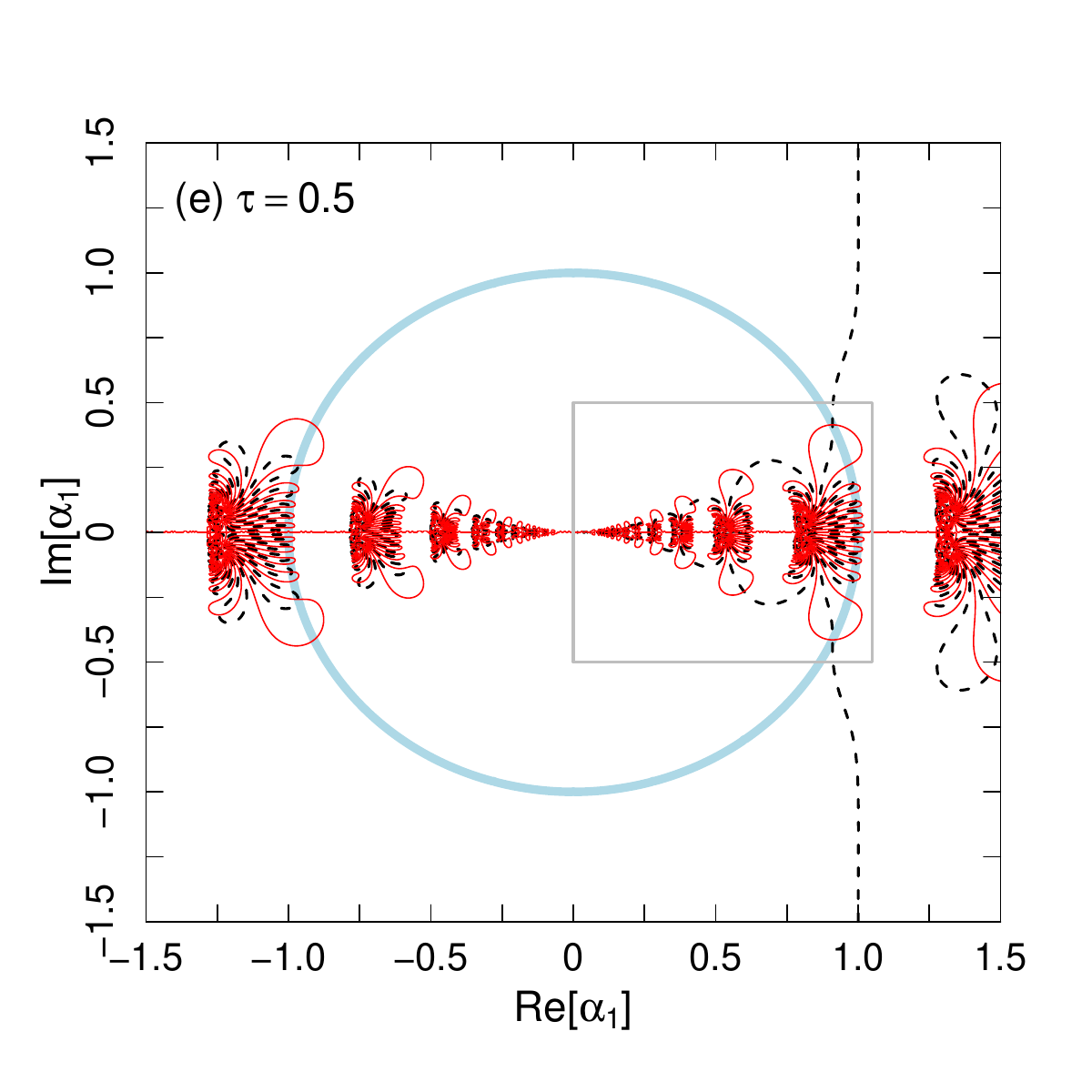} \hspace{-0.5cm}
	\includegraphics[width=6cm]{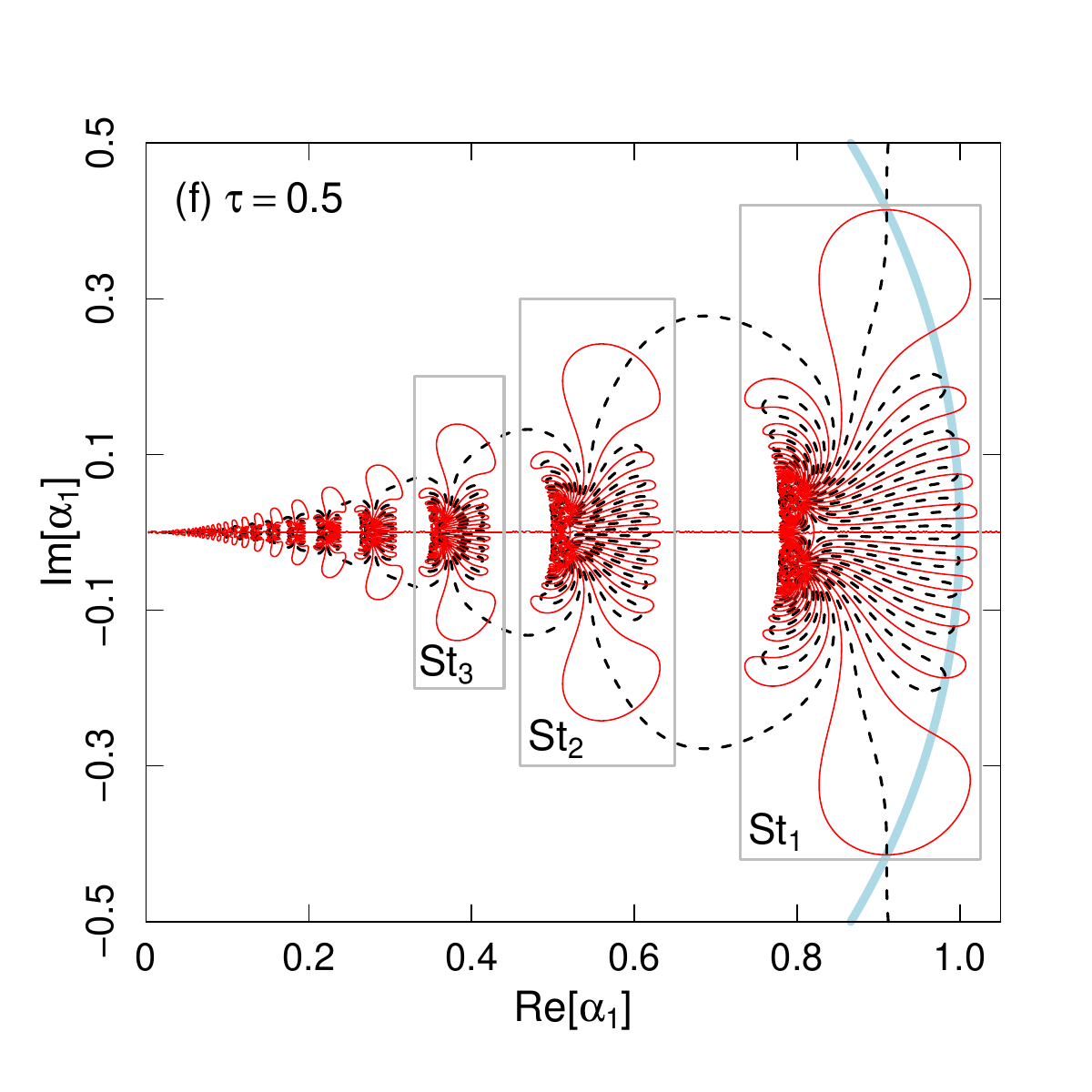}}
\centerline{
	\includegraphics[width=6cm]{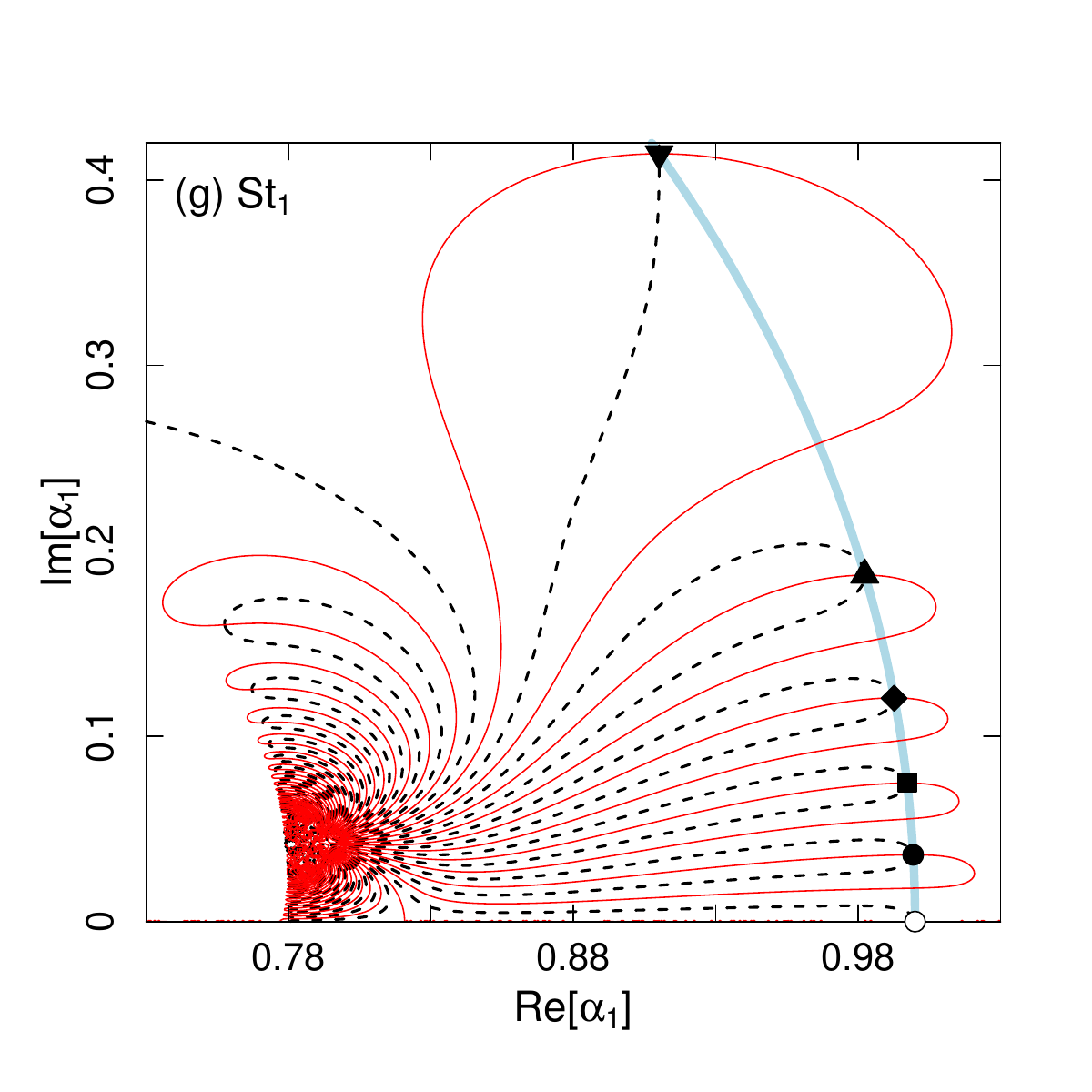} \hspace{-0.5cm}
	\includegraphics[width=6cm]{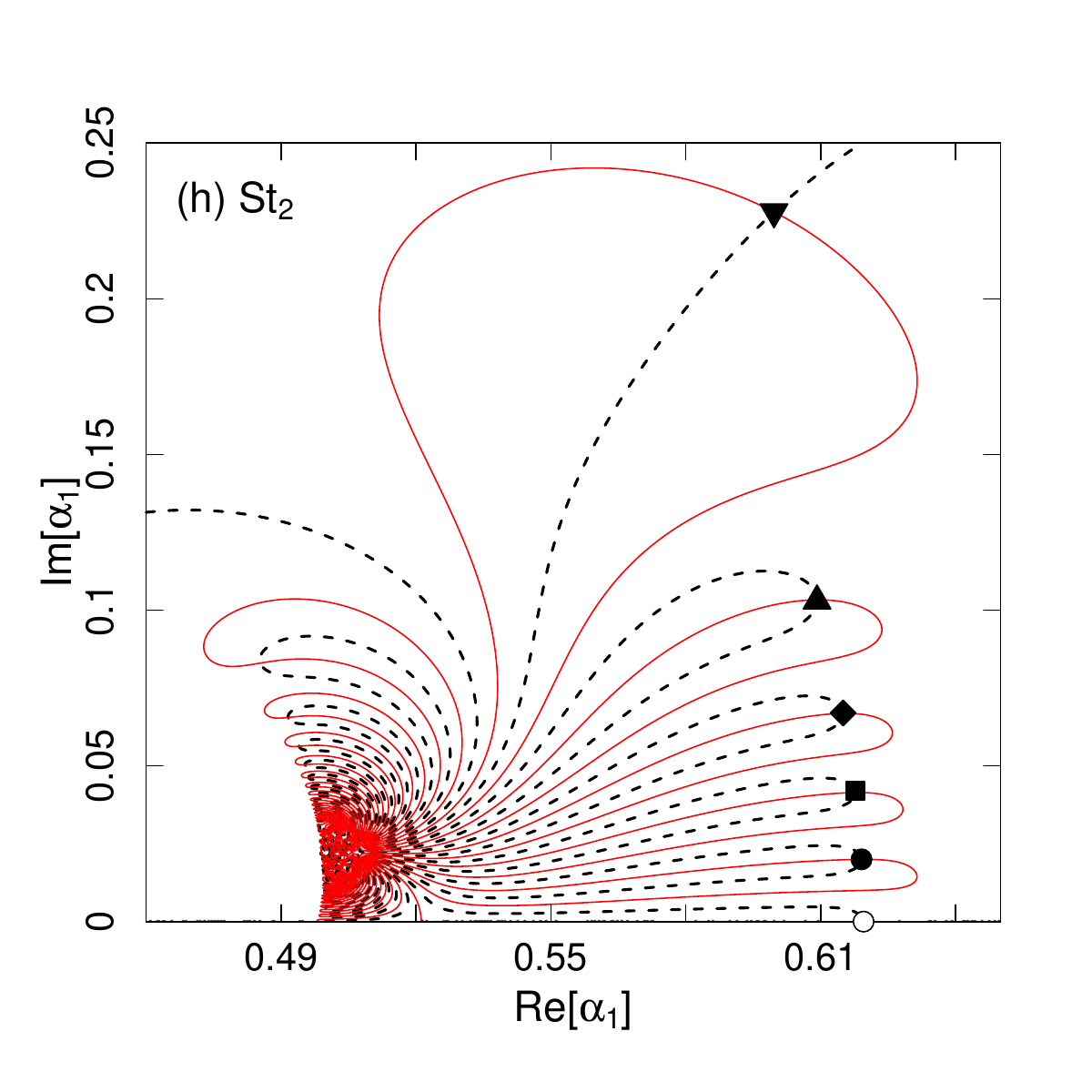} \hspace{-0.5cm}
	\includegraphics[width=6cm]{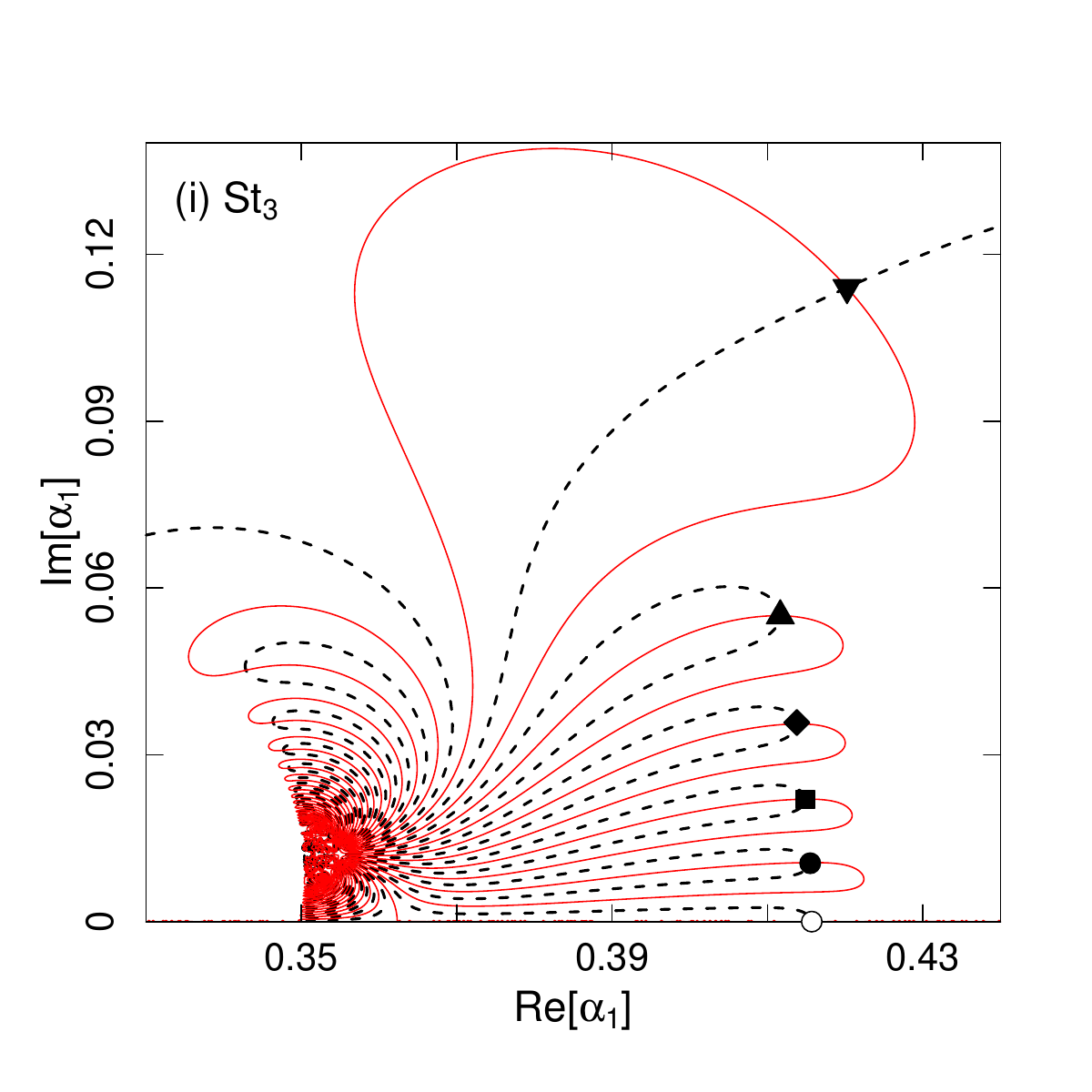}}
	\caption{Contour lines of $f^{\text{\tiny R}}(\alpha_1)=0$ (dashed black lines) and $f^{\text{\tiny I}}(\alpha_1)=0$ (solid red lines) in the $\alpha_1$ complex plane. The thick-solid blue line represents the unit circle. The points where a black curve crosses a red one is a solution of Eq.~\eqref{TranscEq}. Panels (a) to (e) refer to different values of $\tau$: $10^{-5}$, $0.05$, $0.1$, $0.2$, and $0.5$, respectively. Panel (f) shows an amplification of the region marked with a gray square in panel (e) and identifies three structures $\mathrm{St}_1$, $\mathrm{St}_2$, and $\mathrm{St}_3$. At last, panels (g) to (i) amplify each structure seen in panel (f). We plot the curves using the following numerical values: $\lambda=|z_0|^2=|s_0|^2=1$ and $j=5$.}
	\label{fig1}
\end{figure}

\begin{figure}
\centerline{
	\includegraphics[width=9cm]{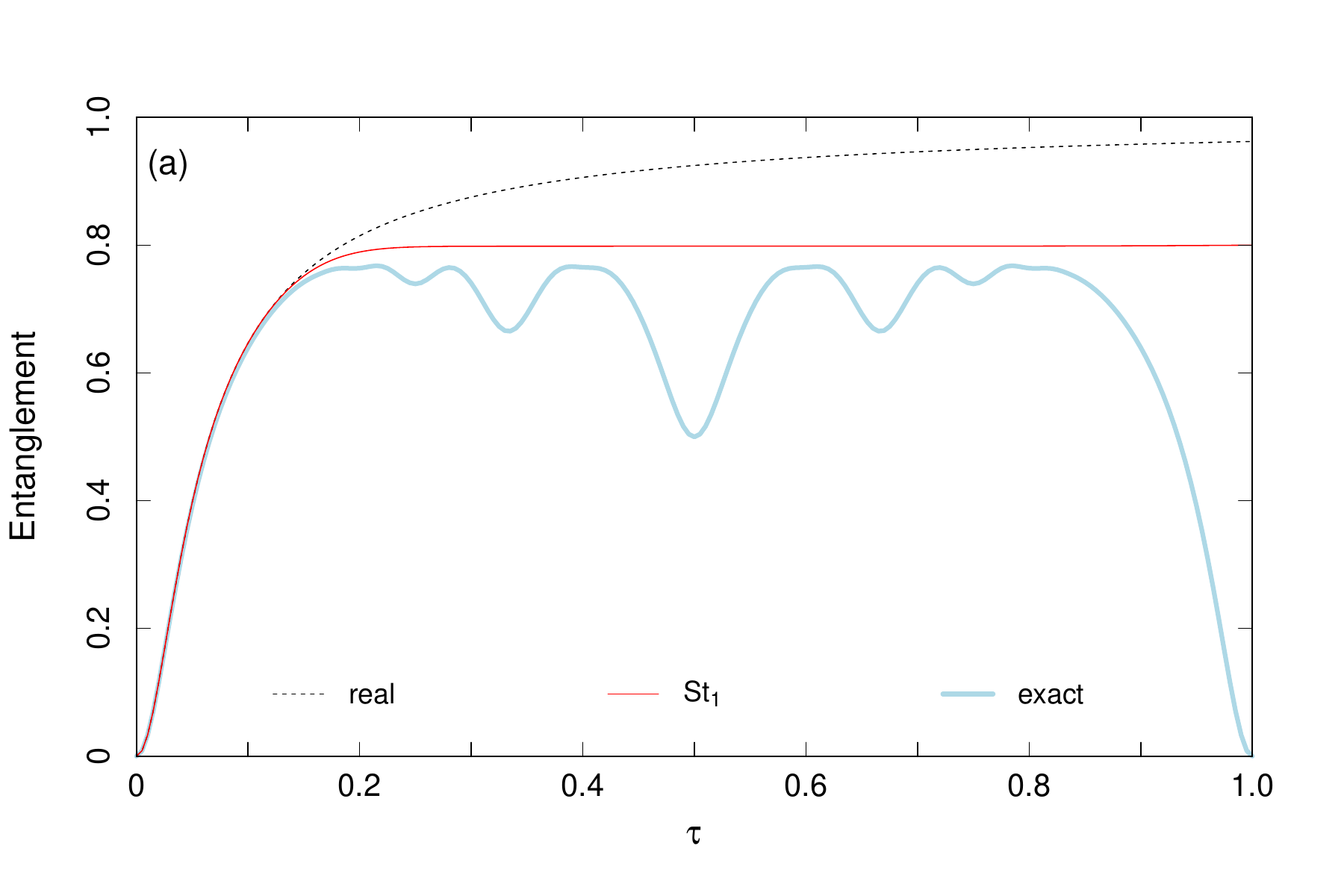} \hspace{-0.5cm}
	\includegraphics[width=9cm]{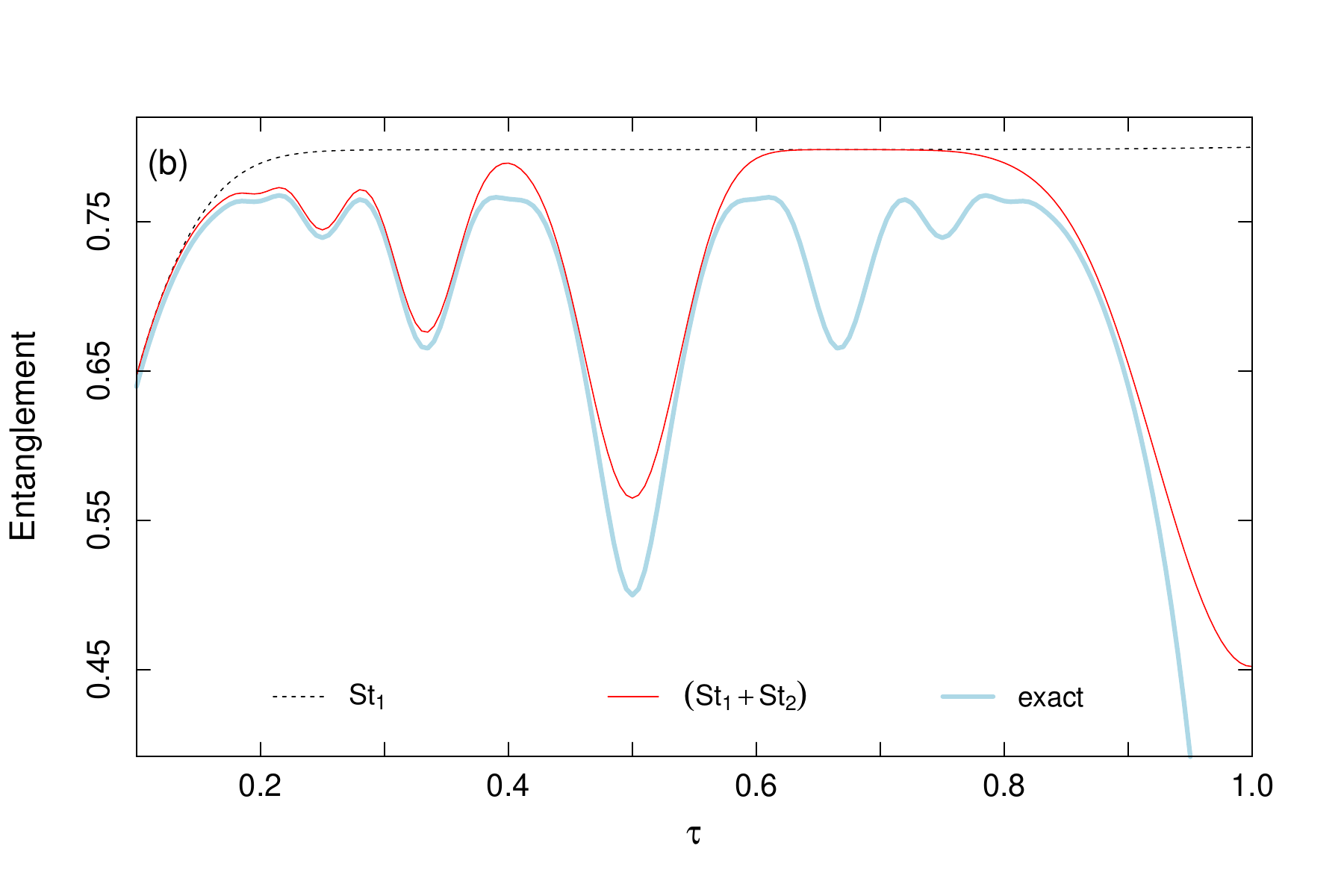}}
\centerline{
	\includegraphics[width=9cm]{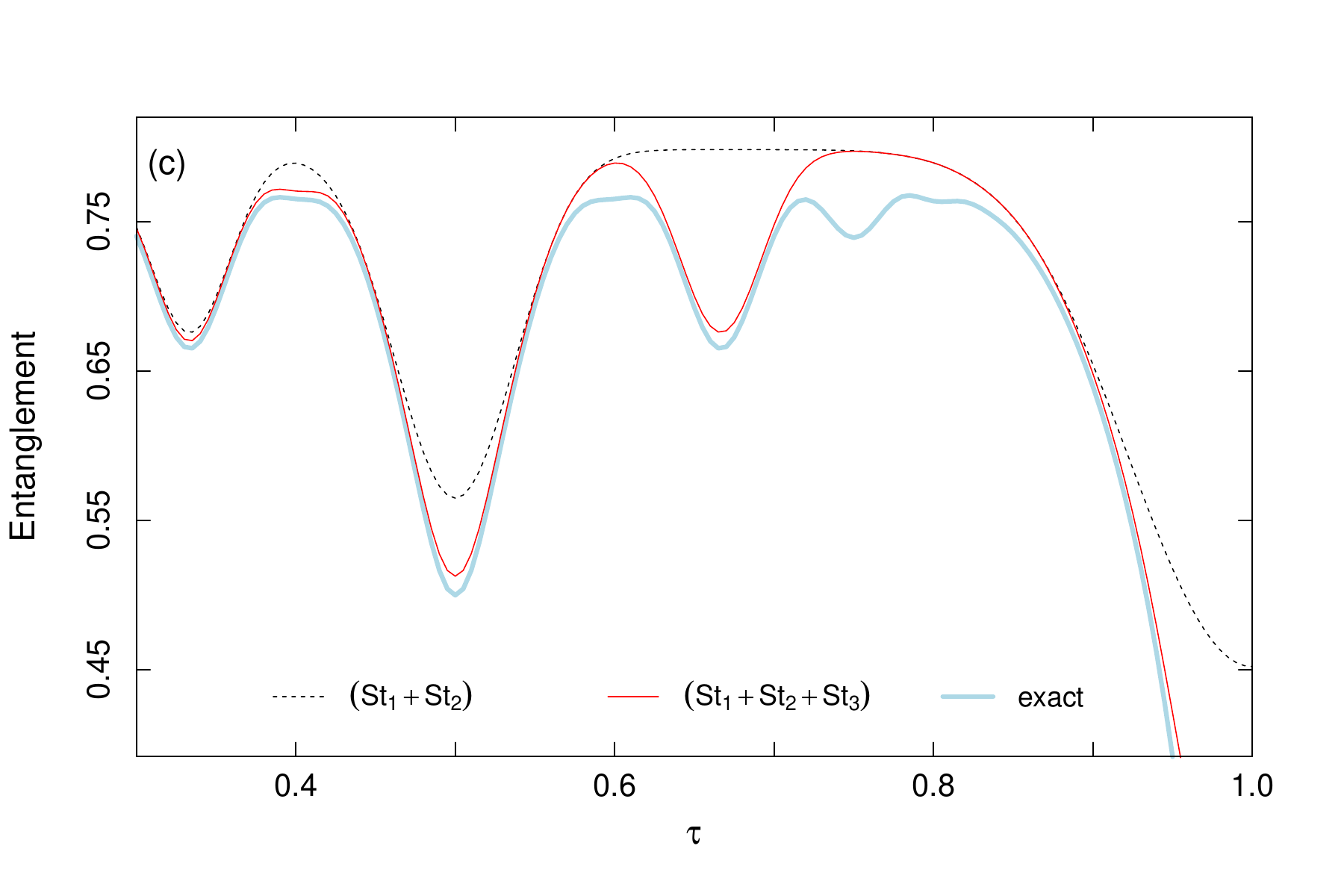} \hspace{-0.5cm}
	\includegraphics[width=9cm]{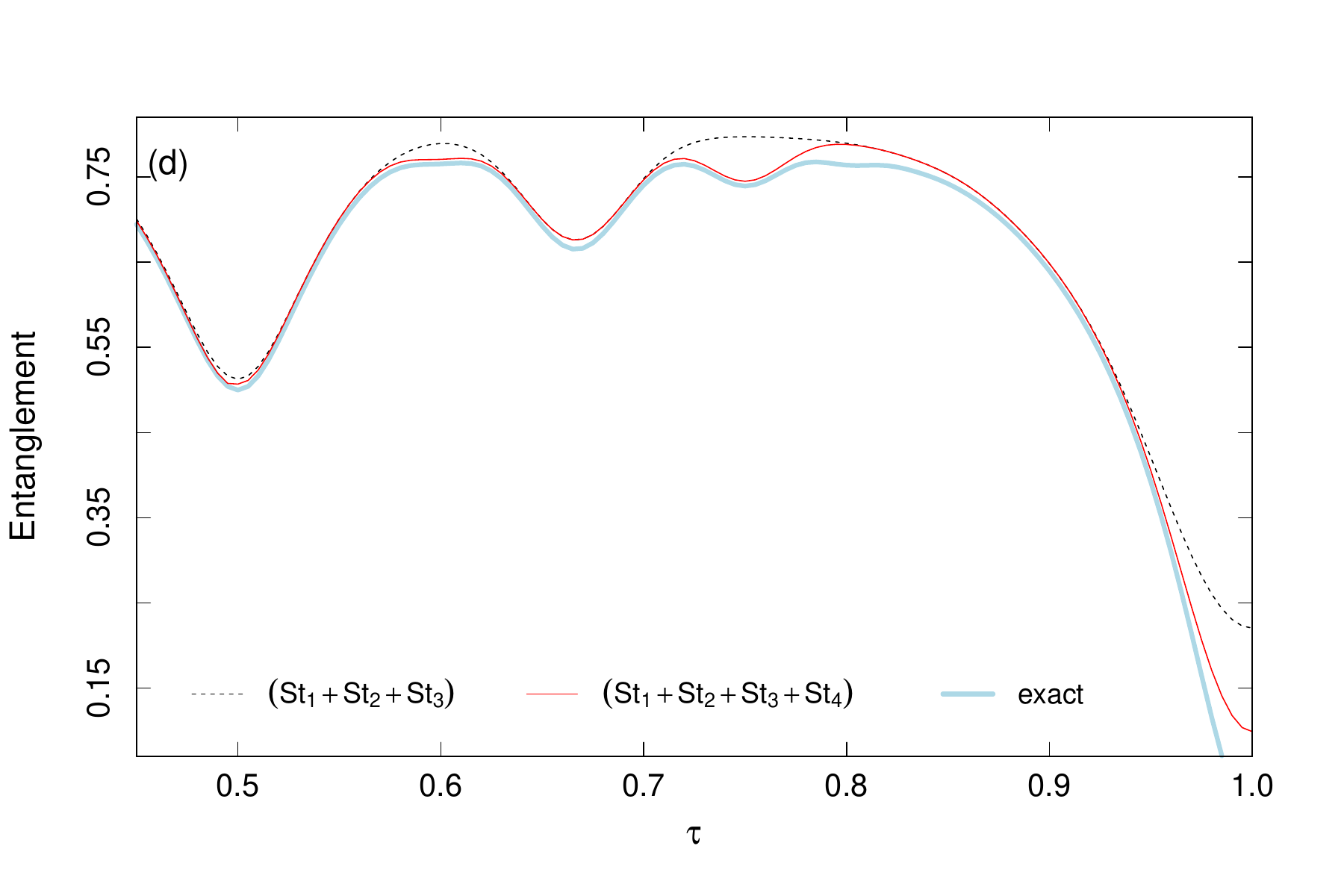}}
\centerline{
	\includegraphics[width=9cm]{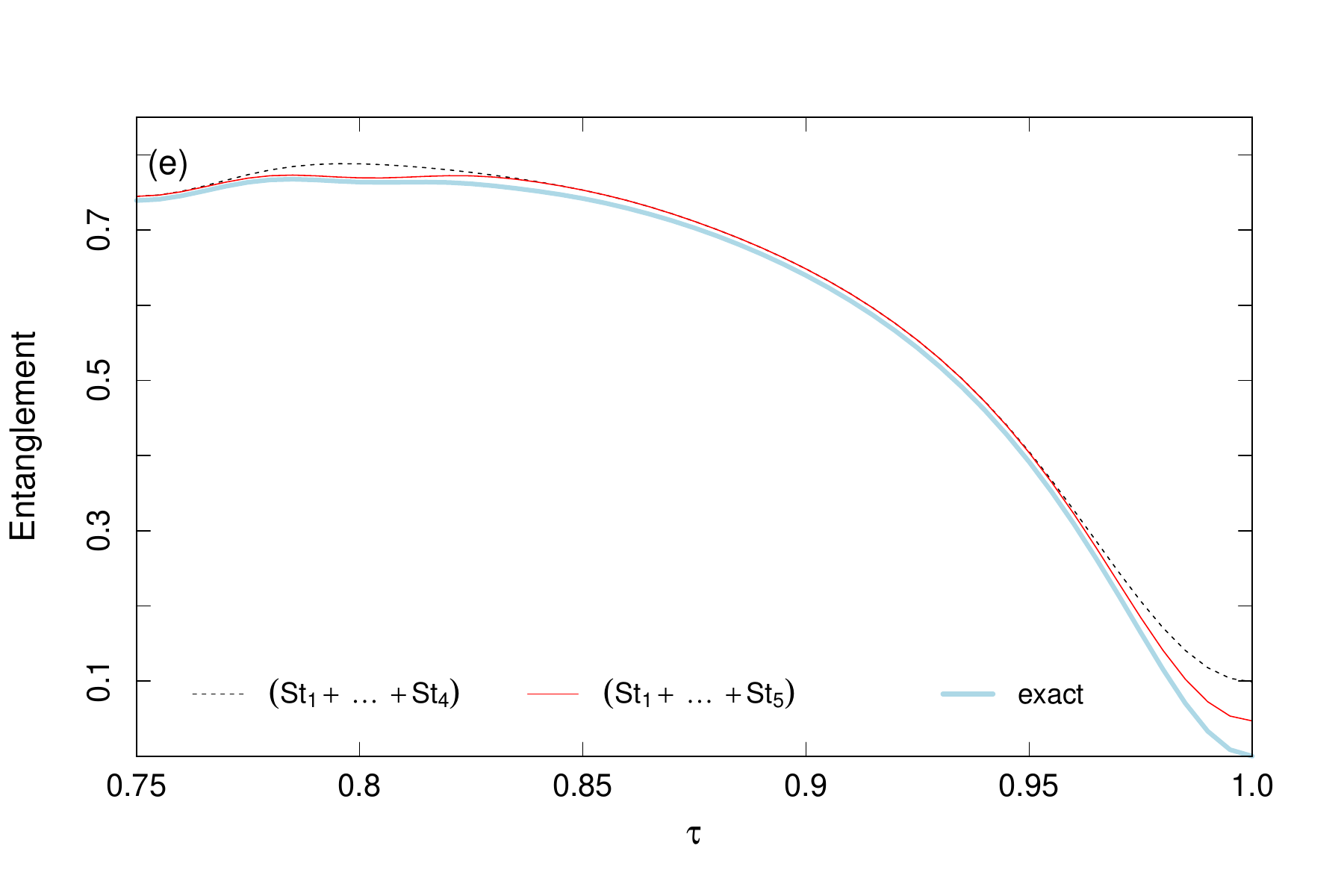} \hspace{-0.5cm}
	\includegraphics[width=9cm]{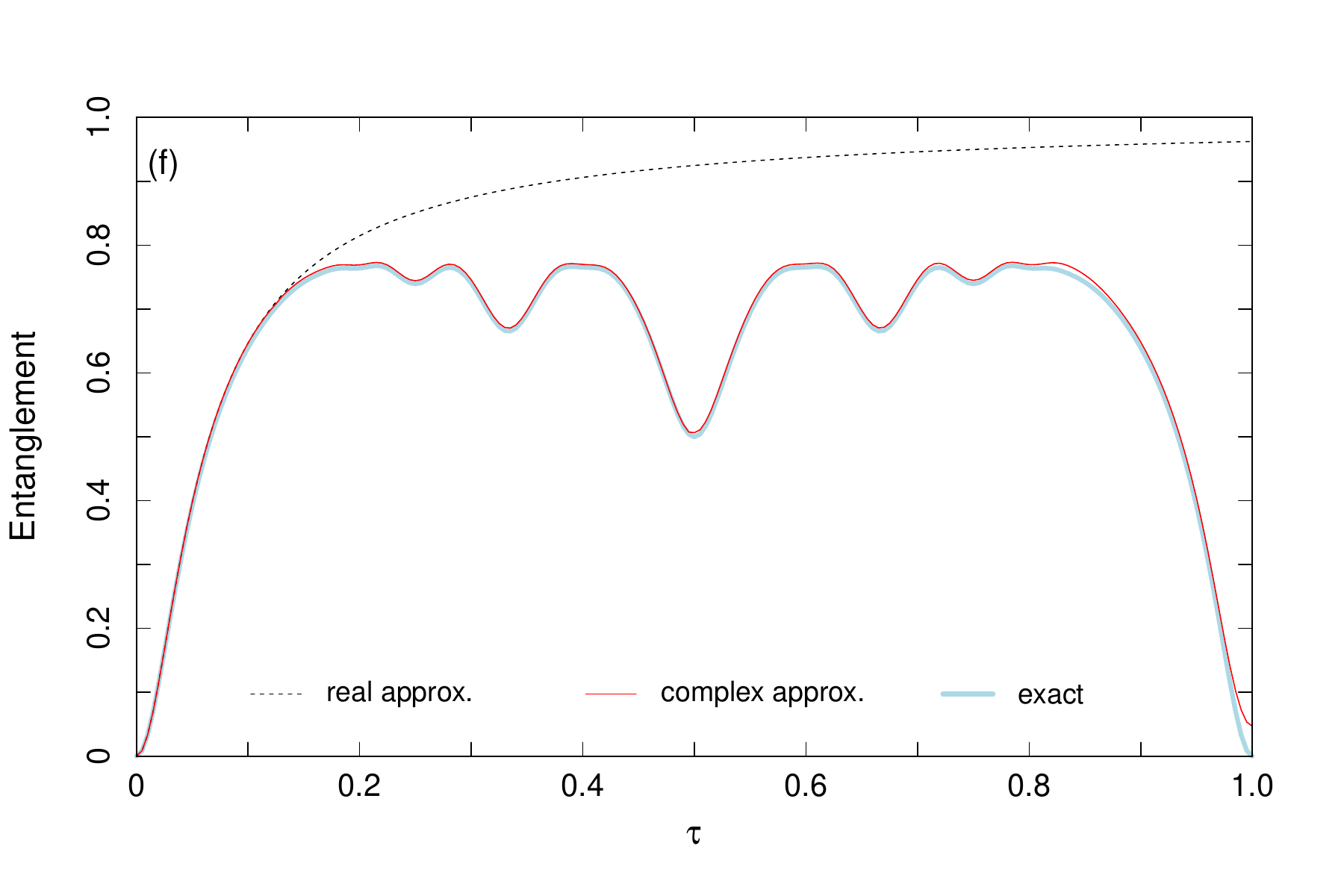}}
\caption{Entanglement as function of time $\tau$. In all panels, the thick solid blue line represents the entirely quantum calculation $E^{\mathrm{q}}_T(\hat{\rho}_0)$ while the other curves show its semiclassical approximations. In panel (a), the dashed black line refers to the semiclassical entanglement~\eqref{Efinal}, including only real trajectories. The solid red line also represents $E^{\mathrm{sc}}_T(\hat{\rho}_0)$, but considering the complex trajectories generated by the roots of $f(\alpha_1)$ belonging to structure $\mathrm{St}_1$. This curve is repeated in panel (b), now as a dashed black line, to be compared with the semiclassical entanglement formula including the trajectories associated with structure $\mathrm{St}_2$ (solid red line). Panels (c)-(e) show the results of the inclusion of the complex trajectories associated with the structures $\mathrm{St}_3$,  $\mathrm{St}_4$, and  $\mathrm{St}_5$, following the same reasoning used from panel (a) to (b). That is, in a given panel, we always reproduce the better semiclassical result of the previous panel for comparison.}
	\label{fig2}
\end{figure}

In Fig.~\ref{fig1}, we plot the contour lines of $f^{\text{\tiny R}}(\alpha_1)=0$ and $f^{\text{\tiny I}}(\alpha_1)=0$ in the $\alpha_1$ complex plane, as dashed black and solid red lines, respectively. For completeness, we also plot the unit circle with a thick solid blue curve. The numerical values of the pertinent parameters are $\lambda=|z_0|^2=|s_0|^2=1$, $j=5$, and $\tau$ is specified in each panel. Notice that the crossings of black and red curves seen in the plots identify the solutions of Eq.~\eqref{TranscEq}. As already discussed before, in Fig.~\ref{fig1}a, we illustrate the only root of $f(z_1)$, appearing at $\alpha_1=1$, for the case $\tau\to0$. This scenario changes as time increases, with many other crossings appearing, as shown by Figs.~\ref{fig1}(b)-\ref{fig1}(e), for $\tau=10^{-5}$, $0.05$, $0.1$, $0.2$, and $0.5$, respectively. We can realize that part of them arises at the origin while another part comes from the region where $\alpha_1\to\infty$, respecting the conclusion~\eqref{sym}. Roughly, these new roots move closer to the unit circle when $\tau$ increases. In Fig.\ref{fig1}(f), we magnify the region inside the gray square shown in the previous panel (e). The idea is to identify three structures ($\mathrm{St}_1$, $\mathrm{St}_2$, and $\mathrm{St}_3$) which exhibit some roots of $f(\alpha_1)$ in an organized manner. We clearly see other structures in panel (f), but we will detail only those three in the present discussion, for shortness.

The structure $\mathrm{St}_1$ is amplified in Fig.~\ref{fig1}(g). There, we can see an open circle indicating the point $\alpha_1=1$, which is the generator of the set of real trajectories. Other five roots, marked with different solid black symbols, are present, and each of them gives origin to a different set of complex trajectories. Notice that a given complex crossing in Fig.~\ref{fig1}(g) generates just one set of complex trajectories, valid specifically for $\tau=0.5$. However, if one continuously increases or decreases the value of $\tau$, it is straightforward to get, using the Newton-Raphson method, a family of four-trajectories sets corresponding to the whole interval $0\le\tau<1$. Thus, in Fig.~\ref{fig2}(a), we illustrate how the inclusion of these five families of complex trajectories [as well as those contemplated by Eq.~\eqref{sym}] improves the accuracy of the semiclassical entanglement formula~\eqref{Efinal}. In this panel (a), the thick solid blue line represents the quantum entanglement $E^{\mathrm{q}}_T(\hat{\rho}_0)$ [see Eq.~\eqref{ETqEx}] while the dashed black curve shows the semiclassical entanglement $E^{\mathrm{sc}}_T(\hat{\rho}_0)$, considering only the {\em real} trajectories. When we include the other contributions mentioned above, we get the solid red curve. This inclusion clearly improves the semiclassical approximation but still does not reproduce the oscillations seen in the quantum (blue) curve.

Figure~\ref{fig1}(h) amplifies structure $\mathrm{St}_2$ and identifies other six roots of Eq.~\eqref{TranscEq}, analogously to Fig.~\ref{fig1}(g). In panel (h), however, all roots give origin to new (families of) sets of complex trajectories. Their inclusion [also considering the contributions implied by Eq.~\eqref{sym}] substantially improves the semiclassical entanglement $E^{\mathrm{sc}}_T(\hat{\rho}_0)$. This result is shown as the solid red line, in Fig.~\ref{fig2}(b). For reference, in the same panel, we include the exact entanglement $E^{\mathrm{q}}_T(\hat{\rho}_0)$ (thick blue curve) and the better semiclassical result of the previous panel (a), now represented by the dashed black curve. Notice how the oscillatory behavior of the quantum entanglement is satisfactorily reproduced by the improved semiclassical formula until $\tau\approx 0.35$. For larger values of $\tau$, however, the red line becomes a rough approximation, indicating that new complex trajectories are welcome.

In Fig.~\ref{fig2}(c), we calculate the semiclassical entanglement $E^{\mathrm{sc}}_T(\hat{\rho}_0)$ considering the sets of complex trajectories associated with the roots seen in structure $\mathrm{St}_3$, shown in Fig.~\ref{fig1}(i). This new result is represented by the solid red line. Again, for comparison, we repeat the thick blue curve in this plot, as well as the better semiclassical result of the previous panel (b), now as a dashed black line. Including those sets of trajectories improves ever more the accuracy of Eq.~\ref{Efinal}. As a last effort to get a better semiclassical result, we consider other two structures ($\mathrm{St}_4$ and $\mathrm{St}_5$) that appear in Fig.~\ref{fig1}(f) at the left side of $\mathrm{St}_3$, but not identified. The effect of these new inclusions is shown in Figs.~\ref{fig2}(d) and~\ref{fig2}(e), in the same way as we presented the previous (a)-(c) panels. Finally, our better semiclassical result, including the complex trajectories associated with the five mentioned structures, is represented in Fig.~\ref{fig2}(f) by the solid red line. It is really impressive how this curve well approaches the exact result (thick blue curve). For comparison, in the same panel, we plot our simplest semiclassical result based on real trajectories with a dashed black line. A last comment should be made. Besides the crossings identified in the above discussion, we can also see others in Fig.~\ref{fig1}. However, their inclusion is either negligible or yields unphysical divergent results. Therefore, we disregarded them.

%%%%%%%%%%%%%%%%%%%%
%%%%%%%%%%%%%%%%%%%%
\section{Final Remarks}
\label{FRsec}

In this work, we have derived a semiclassical formula for the entanglement entropy, written in terms of the trajectories of the corresponding classical description. Our results demonstrate that, as in other bipartite systems, the inclusion of complex trajectories significantly enhances the accuracy of the semiclassical approximation, allowing it to faithfully reproduce the exact quantum dynamics beyond the short-time (Ehrenfest) regime.

The methodology introduced here opens the way for systematic semiclassical studies of more intricate spin–field models, including chaotic and strongly interacting systems. Such extensions are particularly promising for uncovering the classical mechanisms that govern entanglement generation and for clarifying how the degree of chaoticity in phase space influences entanglement growth at longer times.

Finally, we note the connection between our semiclassical formalism and recent proposals linking entanglement dynamics with number theory, such as the use of similar models to encode prime distributions~\cite{southier2023,jonas2024}. In the future, we plan to explore  our semiclassical approach to investigate the role of complex trajectories in determining primes. This idea is inspired by a work by Berry and Keating~\cite{BerryKeating}, where the energy spectrum of a given Hamiltonian is semiclassically evaluated by summing contributions of periodic orbits, and the achieved expressions are proved to be similar to the distribution of the zeros of the Riemann Zeta function. In our case, we will investigate whether entanglement entropy, written in terms of complex trajectories, leaves signatures related to the distribution of prime numbers. Such studies could further enrich the interplay between semiclassical physics, entanglement theory, and mathematical structures underlying quantum systems.

%%%%%%%%%%%%%%%%%%%%%%%%%
\section{ACKNOWLEDGMENTS}

A.D.R. thanks the financial support from the National Institute for Science and Technology of Quantum Information (CNPq, INCT-IQ 465469/2014-0). He is also grateful to the Department of Physics at the University of Connecticut for their hospitality during his stay in Storrs, where part of this work was carried out.

%%%%%%%%%%%%%%%%%%%%
%%%%%%%%%%%%%%%%%%%%
\appendix
\section{Derivatives of the action and the tangent matrix \label{ap1}} 

By differentiating the first line of Eq.~\eqref{dS}, we get
\begin{equation}
\mathbf{B}'' \, \delta\mathbf{u}'' =
\mathbf{S}^{+}_{\mathrm{vu}} \, \delta\mathbf{u}' + 
\left[ \mathbf{S}^{+}_{\mathrm{vv}} + {u''_{\mbox{\tiny B}}}^2 \mathbf{A}'' \right]\delta \mathbf{v}'' 
\quad\mathrm{and}\quad
\mathbf{B}' \, \delta\mathbf{v}' =
\mathbf{S}^{+}_{\mathrm{uv}} \, \delta\mathbf{v}'' + 
\left[ \mathbf{S}^{+}_{\mathrm{uu}} + {v'_{\mbox{\tiny B}}}^2\mathbf{A}' \right]\delta \mathbf{u}' ,
\label{d2+}
\end{equation}
where we define the colunm vectors $\delta \mathbf{u}\equiv(\delta u_{\mbox{\tiny A}},\delta u_{\mbox{\tiny B}})$ and $\delta \mathbf{v} \equiv (\delta v_{\mbox{\tiny A}},\delta v_{\mbox{\tiny B}})$, and the matrices
\begin{eqnarray}
\mathbf A \equiv 
\left(\begin{array}{cc}
0&0 \\ 
0& \frac{-2ij\hbar}{(1+u_{\mbox{\tiny B}} v_{\mbox{\tiny B}})^2}
\end{array}\right), 
\quad
\mathbf B \equiv 
\left(\begin{array}{cc}
-i\hbar&0 \\ 
0& \frac{-2ij\hbar}{(1+u_{\mbox{\tiny B}} v_{\mbox{\tiny B}})^2}
\end{array}\right), 
\quad\mathrm{and}\quad
\mathbf{S}^{+}_{\mathrm{uv}} \equiv
\left(\begin{array}{cc}
\frac{\partial^2\mathcal S_+}{\partial u'_{\mbox{\tiny A}}\partial v''_{\mbox{\tiny A}}} & 
\frac{\partial^2\mathcal S_+}{\partial u'_{\mbox{\tiny A}}\partial v''_{\mbox{\tiny B}}} \\
\frac{\partial^2\mathcal S_+}{\partial u'_{\mbox{\tiny B}}\partial v''_{\mbox{\tiny A}}} & 
\frac{\partial^2\mathcal S_+}{\partial u'_{\mbox{\tiny B}}\partial v''_{\mbox{\tiny B}}} 
\end{array}\right).
\nonumber
\end{eqnarray}
In addition, we use prime (double prime) to denote initial (final) time, and the other matrices $\mathbf{S}^{+}_{\mathrm{uu}}$, $\mathbf{S}^{+}_{\mathrm{vu}}$, and $\mathbf{S}^{+}_{\mathrm{vv}}$ are defined in analogy to $\mathbf{S}^{+}_{\mathrm{uv}}$. Now, we rearrange Eq.~(\ref{d2+}) to get the same form of Eq.~\eqref{Mtangent}. We find the following expressions
\begin{equation}
\begin{array}{lll}
\mathbf{M}_{\mathrm{uu}}&=&
\big[ \mathbf{B}'' \big]^{-1} \left\{\mathbf{S}^{+}_{\mathrm{vu}}
-\big[ \mathbf{S}^{+}_{\mathrm{vv}}+{u''_{\mbox{\tiny B}}}^2 \mathbf{A}''\big]
\big[\mathbf{S}^{+}_{\mathrm{uv}}\big]^{-1}
\big[ \mathbf{S}^{+}_{\mathrm{uu}}+{v'_{\mbox{\tiny B}}}^2 \mathbf{A}'\big]\right\},\\
\mathbf{M}_{\mathrm{uv}}&=&
\big[\mathbf{B}''\big]^{-1}
\big[ \mathbf{S}^{+}_{\mathrm{vv}}+{u''_{\mbox{\tiny B}}}^2 \mathbf{A}''\big]
\big[\mathbf{S}^{+}_{\mathrm{uv}}\big]^{-1}
\mathbf{B}',\\
\mathbf{M}_{\mathrm{vu}}&=&
-\big[\mathbf{S}^{+}_{\mathrm{uv}}\big]^{-1}
\big[ \mathbf{S}^{+}_{\mathrm{uu}}+{v'_{\mbox{\tiny B}}}^2 \mathbf{A}'\big],\\
\mathbf{M}_{\mathrm{vv}}&=&
\big[\mathbf{S}^{+}_{\mathrm{uv}}\big]^{-1}\mathbf{B}' .
\end{array}\label{MS+}
\end{equation}
At last, we invert them to find
\begin{equation}
\begin{array}{lll}
\mathbf{S}^{+}_{\mathrm{uu}} &=& 
-\mathbf{B}' \big[\mathbf{M}_{\mathrm{vv}}\big]^{-1}
\mathbf{M}_{\mathrm{vu}}- {v'_{\mbox{\tiny B}}}^2 \mathbf{A}', \\
\mathbf{S}^{+}_{\mathrm{uv}}&=& 
\mathbf{B}' \big[\mathbf{M}_{\mathrm{vv}} \big]^{-1} ,\\
\mathbf{S}^{+}_{\mathrm{vu}}&=& 
\mathbf{B}'' \left\{\mathbf{M}_{\mathrm{uu}}-
\mathbf{M}_{\mathrm{uv}} \big[\mathbf{M}_{\mathrm{vv}}\big]^{-1}
\mathbf{M}_{\mathrm{vu}}
\right\}],\\
\mathbf{S}^{+}_{\mathrm{vv}}&=& 
\mathbf{B}''\mathbf{M}_{\mathrm{uv}}\big[\mathbf{M}_{\mathrm{vv}}\big]^{-1}-
{u''_{\mbox{\tiny B}}}^2\mathbf{A}''.
\end{array} \label{SM+}
\end{equation}

We will now find analogous relations for the backward propagator. We first differentiate the second line of Eq.~(\ref{dS}), finding
\begin{equation}
\mathbf{B}' \, \delta\mathbf{u}' =
\mathbf{S}^{-}_{\mathrm{vu}} \, \delta\mathbf{u}'' + 
\left[ \mathbf{S}^{-}_{\mathrm{vv}} + {u'_{\mbox{\tiny B}}}^2 \mathbf{A}' \right]\delta \mathbf{v}' 
\quad\mathrm{and}\quad
\mathbf{B}'' \, \delta\mathbf{v}'' =
\mathbf{S}^{-}_{\mathrm{uv}} \, \delta\mathbf{v}' + 
\left[ \mathbf{S}^{-}_{\mathrm{uu}} + {v''_{\mbox{\tiny B}}}^2\mathbf{A}'' \right]\delta \mathbf{u}'' .
\label{d2-}
\end{equation}
Then, we manipulate them to get
\begin{equation}
\begin{array}{lll}
\mathbf{M}_{\mathrm{uu}} &=&
\big[\mathbf{S}^{-}_{\mathrm{vu}}\big]^{-1}\mathbf{B}',\\
\mathbf{M}_{\mathrm{uv}} &=&
-\big[\mathbf{S}^{-}_{\mathrm{vu}}\big]^{-1} 
\big[ \mathbf{S}^{-}_{\mathrm{vv}} + {u'_{\mbox{\tiny B}}}^2 {\mathbf{A}}' \big] ,\\
\mathbf{M}_{\mathrm{vu}} &=&
\big[ \mathbf{B}''\big]^{-1} 
\big[ \mathbf{S}^{-}_{\mathrm{uu}} + {v''_{\mbox{\tiny B}}}^2 {\mathbf{A}}'' \big] 
\big[\mathbf{S}^{-}_{\mathrm{vu}}\big]^{-1}
\mathbf{B}',\\
\mathbf{M}_{\mathrm{vv}} &=& 
\big[\mathbf{B}''\big]^{-1}\left\{
\mathbf{S}^{-}_{\mathrm{uv}} - 
\big[ \mathbf{S}^{-}_{\mathrm{uu}} + {v''_{\mbox{\tiny B}}}^2 {\mathbf{A}}'' \big] 
\big[\mathbf{S}^{-}_{\mathrm{vu}}\big]^{-1}
\big[ \mathbf{S}^{-}_{\mathrm{vv}} + {u'_{\mbox{\tiny B}}}^2 {\mathbf{A}}' \big] 
\right\},
\end{array}\label{MS-}
\end{equation}
and
\begin{equation}
\begin{array}{lll}
\mathbf{S}^{-}_{\mathrm{uu}} &=& 
\mathbf{B}''\mathbf{M}_{\mathrm{vu}}\big[\mathbf{M}_{\mathrm{uu}}\big]^{-1}-
{v''_{\mbox{\tiny B}}}^2   \mathbf{A}'',\\
\mathbf{S}^{-}_{\mathrm{uv}} &=&
\mathbf{B}'' \left\{
\mathbf{M}_{\mathrm{vv}}-\mathbf{M}_{\mathrm{vu}}\big[\mathbf{M}_{\mathrm{uu}}\big]^{-1}
\mathbf{M}_{\mathrm{uv}}
\right\} ,\\
\mathbf{S}^{-}_{\mathrm{vu}} &=& \mathbf{B}' \big[\mathbf{M}_{\mathrm{uu}}\big]^{-1} ,\\
\mathbf{S}^{-}_{\mathrm{vv}} &=&-\mathbf{B}'
\big[\mathbf{M}_{\mathrm{uu}}\big]^{-1}\mathbf{M}_{\mathrm{uv}}-
{u'_{\mbox{\tiny B}}}^2   \mathbf{A}'.
\end{array} \label{SM-}
\end{equation}

Notice that Eqs. (\ref{MS+}), (\ref{SM+}), (\ref{MS-}), and (\ref{SM-}) establish the connection between second derivatives of the complex classical action $S_{\pm}$ [Eq.~\eqref{S}], for both forward and backward propagators, with the elements of the stability matrix~\eqref{Mtangent}. 

%%%%%%%%%%%%%%%%%%%%
%%%%%%%%%%%%%%%%%%%%
\bibliography{SCSCE_subm}
\end{document}